\documentclass[%
 aip,amsmath,amssymb,floatfix, reprint,
 jcp
]{revtex4-2}

\usepackage{dsfont}
\usepackage{svg,bbm}
\usepackage{graphicx}
\usepackage{dcolumn}
\usepackage{physics}
\usepackage{amsmath}
\usepackage{amsfonts}
\usepackage{bbold}
\usepackage{float}

\usepackage{hyperref}
\hypersetup{
    colorlinks=true,
	citecolor=blue,
    linkcolor=red,
	urlcolor = blue
}

\setcitestyle{super}

\newcommand*{\citen}[1]{%
  \begingroup
    \romannumeral-`\x 
    \setcitestyle{numbers}%
    \cite{#1}%
  \endgroup   
}

\usepackage{bm}
\usepackage[mathlines]{lineno}

\newcommand{\tauL}{\tau^{\mathcal{L}}_{\rm m}}

\preprint{APS/123-QED}

\begin{document}
\title{Extracting Dynamical Maps of Non-Markovian Open Quantum Systems}%
\author{David J. Strachan}
\email{david.strachan@bristol.ac.uk}
\affiliation{\textit{H. H. Wills Physics Laboratory, University of Bristol, Bristol BS8 1TL, United Kingdom}}
\author{Archak Purkayastha}
\email{archak.p@phy.iith.ac.in}
\affiliation{\textrm{Department of Physics, Indian Institute of Technology, Hyderabad 502284, India}}
\affiliation{Centre for complex quantum systems, Aarhus University, Nordre Ringgade 1, 8000 Aarhus C, Denmark}
\author{Stephen R. Clark}
\email{stephen.clark@bristol.ac.uk}
\affiliation{\textit{H. H. Wills Physics Laboratory, University of Bristol, Bristol BS8 1TL, United Kingdom}}
\date{\today}    
\begin{abstract}
The most general description of quantum evolution up to a time $\tau$ is a completely positive tracing preserving map known as a {\em dynamical map} $\hat{\Lambda}(\tau)$. Here we consider $\hat{\Lambda}(\tau)$ arising from suddenly coupling a system to one or more thermal baths with a strength that is neither weak nor strong. Given no clear separation of characteristic system/bath time scales $\hat{\Lambda}(\tau)$ is generically expected to be non-Markovian, however we do assume the ensuing dynamics has a unique steady state implying the baths possess a finite memory time $\tau_{\rm m}$. By combining several techniques within a tensor network framework we directly and accurately extract $\hat{\Lambda}(\tau)$ for a small number of interacting fermionic modes coupled to infinite non-interacting Fermi baths. First, we use an orthogonal polynomial mapping and thermofield doubling to arrive at a purified chain representation of the baths whose length directly equates to a time over which the dynamics of the infinite baths is faithfully captured. Second, we employ the Choi-Jamiolkowski isomorphism so that $\hat{\Lambda}(\tau)$ can be fully reconstructed from a single pure state calculation of the unitary dynamics of the system, bath and their replica auxillary modes up to time $\tau$. From $\hat{\Lambda}(\tau)$ we also compute the time local propagator $\hat{\mathcal{L}}(\tau)$. By examining the convergence with $\tau$ of the instantaneous fixed points of these objects we establish their respective memory times $\tau^{\Lambda}_{\rm m}$ and $\tauL$. Beyond these times, the  propagator $\hat{\mathcal{L}}(\tau)$ and dynamical map $\hat{\Lambda}(\tau)$ accurately describe all the subsequent long-time relaxation dynamics up to stationarity. These timescales form a hierarchy $\tauL \leq \tau^{\Lambda}_{\rm m} \leq \tau_{\textrm{re}}$, where $\tau_{\textrm{re}}$ is a characteristic relaxation time of the dynamics. Our numerical examples of interacting spinless Fermi chains and the single impurity Anderson model demonstrate regimes where $\tau_{\textrm{re}} \gg \tau_{\rm m}$, where our approach can offer a significant speedup in determining the stationary state compared to directly simulating the long-time limit. Our results also show that having access to $\hat{\Lambda}(\tau)$ affords a number of insightful analyses of the open system thus far not commonly exploited.
\end{abstract}
\maketitle

\section{Introduction}

The problem of theoretically obtaining transient relaxation dynamics and steady states of open many-body quantum systems, both in and out of equilibrium is a notoriously challenging problem. Yet, a better understanding of dissipation and decoherence is crucial for understanding real-world quantum systems. The recent developments in nanoscale devices have motivated interest in technologies that harness quantum effects to greatly outperform their classical analogues \cite{nielsen2001quantum,josefsson2018quantum,ronzani2018tunable,mosso2019thermal}, while the existence of non-trivial quantum effects in biological systems is being increasingly investigated \cite{lambert2013quantum,van2000photosynthetic,may2023charge}. Existing approaches to modelling open quantum systems often use perturbative arguments, requiring a clear separation of energy or time scales. A popular approach is the master equation formalism, based on the assumptions of weak system-bath coupling and the Markov approximation. However, their application is limited and can lead to drastically incorrect predictions for composite systems, where local descriptions of dissipation violate thermodynamic laws \cite{Levy_2014,Stockburger_2016} while global descriptions \cite{PhysRevE.76.031115} fail to capture the coherences necessary to describe non-equilibrium steady states, even in the limit of weak coupling \cite{Mitchison_2018}. 

Alternatively, non-equilibrium Green functions \cite{stefanucci_vanleeuwen_2013} can be used to model energy transport under strong system-reservoir coupling, but interactions within the systems must be treated perturbatively \cite{wang2014nonequilibrium,talarico2020study,stefanucci2013nonequilibrium}. To proceed with less severe approximations, one must consider the environment more explicitly, a task that appears formidable due to the many, possibly infinite environmental degrees of freedom. However, the true state of a combined system-bath setup is often characterised by an amount of information much smaller than the maximum capacity of the corresponding Hilbert space. This is exploited in the time-evolving matrix product operator (TEMPO) formalism where the effect of the environment on the system is described by Feynman-Vernon influence functionals as matrix-product states in the temporal domain \cite{strathearn2018efficient,PhysRevB.107.195101,FEYNMAN1963118,PhysRevX.11.021040} or more generally a process tensor\cite{Cygorek2022,Pollock2018,Fux2023,Cygorek2024a}. Alternatively, one can use the time derivative hierarchical equations of motion (HEOM) approach \cite{10.1063/5.0011599}. 
In this work, we follow the approach of Kohn and Santoro~\cite{kohn2021quenching} and describe both the system and environment as a single matrix product state which is then time evolved directly. This is motivated by the fact that many open-system Hamiltonians can be mapped onto 1D structures containing only nearest-neighbour interactions \cite{eisert2010colloquium}. As such 1D tensor networks like matrix product states (MPSs) can often be used to accurately reproduce the dynamics of the system-environment setup by compressing the time-evolved state up to a finite bond-dimension $\chi$, defining the dominant numerical error. A crucial result of these 1D chain mappings is the fact that even in the absence of compression, the size of the bath required scales almost linearly with the evolution time required \cite{PhysRevB.92.155126,PhysRevLett.115.130401,Woods_2016}. Therefore, the finite time evolution of a system coupled to an infinite bath can be modelled exactly with a finite bath. 

This observation is exploited in the Periodically Refreshed Baths (PReB) formalism~\cite{Purkayastha_2021} which allows full reconstruction of the open-system dynamics up to time $\tau$ provided there exists a unique steady state. It works by recursively evolving up to a much smaller time $\tau_{m}^{\Lambda}$ associated with the memory of the bath correlations, before tracing out the bath and replacing it with one in its initial product state. This bares much similarity to a collisional model of an open system~\cite{PhysRev.129.1880,PhysRevLett.88.097905,PhysRevA.65.042105,Campbell_2021}, where the bath is represented by a reservoir of identically prepared ancilla each of which interacts with the system once for a short time before being traced out. Typically the ancilla are single sites, whereas in the PReB approach finite-sized chains are used which allows for significant non-Markovian effects and more accurate modelling of generic bath spectral functions. Performing a direct single evolution of the system + bath evolution using tensor networks often results in an intractably large bond dimension in the long-time limit owing to the generic dynamical growth of entanglement. This makes the stationary properties hard to evaluate. The main advantage of PReB is that the entanglement growth within the baths is removed with each refreshment, limiting the growth of $\chi$. 

Motivated by PReB, in this paper we investigate the combination of Thermofield doubling~\cite{takahashi1996thermo} and the Choi-Jamiolkowski isomorphism~\cite{CHOI1975285,JAMIOLKOWSKI1972275} to extract the effect of one PReB cycle on a fermionic system as a dynamical map. Assuming the PReB cycle has evolved well beyond $\tau_{m}^{\Lambda}$ the steady state and all transient dynamics from any initial state can then be extracted from a single simulation of an enlarged system described by a pure state. Additionally, we calculate the time local propagator $\frac{\partial \hat{\rho}}{\partial t} = \hat{\mathcal{L}}(t)[\hat{\rho}(t)]$ and also extract the steady state when it first becomes invariant. This is directly related to the memory kernel in the Nakajima-Zwanzig equation,  $\frac{\partial \hat{\rho}}{\partial t} = -i\int_{t_{0}}^{t}ds\hat{\mathcal{K}}(t,s)\hat{\rho}(s)$, via a fixed point relation \cite{PhysRevB.104.155407}. A key point here is that $\hat{\mathcal{K}}(t,s)$ is often highly nontrivial and intractable to calculate \cite{chruściński2022dynamicalmapsmarkovianregime,10.1063/1.1731409,10.1143/PTP.20.948}. There have been two notable branches of work on calculating memory kernels, the first of which involves numerically solving a Volterra equation of the second kind \cite{10.1063/1.1624830,10.1063/1.2218342,PhysRevB.84.075150,Cohen_2013} and another is based on the transfer matrix approach \cite{PhysRevLett.112.110401,Pollock_2018,10.21468/SciPostPhys.13.2.027}. Here, we provide a method to numerically extract $\hat{\mathcal{L}}(t)$, which gives an equivalent description to $\hat{\mathcal{K}}(t,s)$, and $\hat{\Lambda}(t)$, allowing us direct access to these rich objects. 

The structure of the paper is as follows. In Sec.~\ref{sec:nonmarkov} we outline the general description of open quantum systems via dynamical maps $\hat{\Lambda}(t)$ as well as the time local propagator $\hat{\mathcal{L}}(t)$, culminating in the two approaches for computing the transient relaxation dynamics and stationary properties. This is followed in Sec.~\ref{sec:baths} where we introduce the thermofield purification and the orthogonal polynomial chain mapping of an infinite non-interacting fermionic bath. In Sec.~\ref{sec:dynamical} we introduce the Choi-Jamiolkowski isomorphism, outline the modifications needed to apply this to a fermionic system and discuss solving the dynamics for non-interacting systems and interacting ones using matrix product states. Building on this Sec.~\ref{sec:results} reports the main results of this work demonstrating the extraction of $\hat{\Lambda}(t)$ and $\hat{\mathcal{L}}(t)$ for a small spinless Fermi chain both with and without interactions and the single impurity Anderson model. In Sec.~\ref{sec:conclusions} we conclude and provide an outlook for future work. 

\begin{figure}[t]
  \centering
  \includegraphics[width=0.35\textwidth]{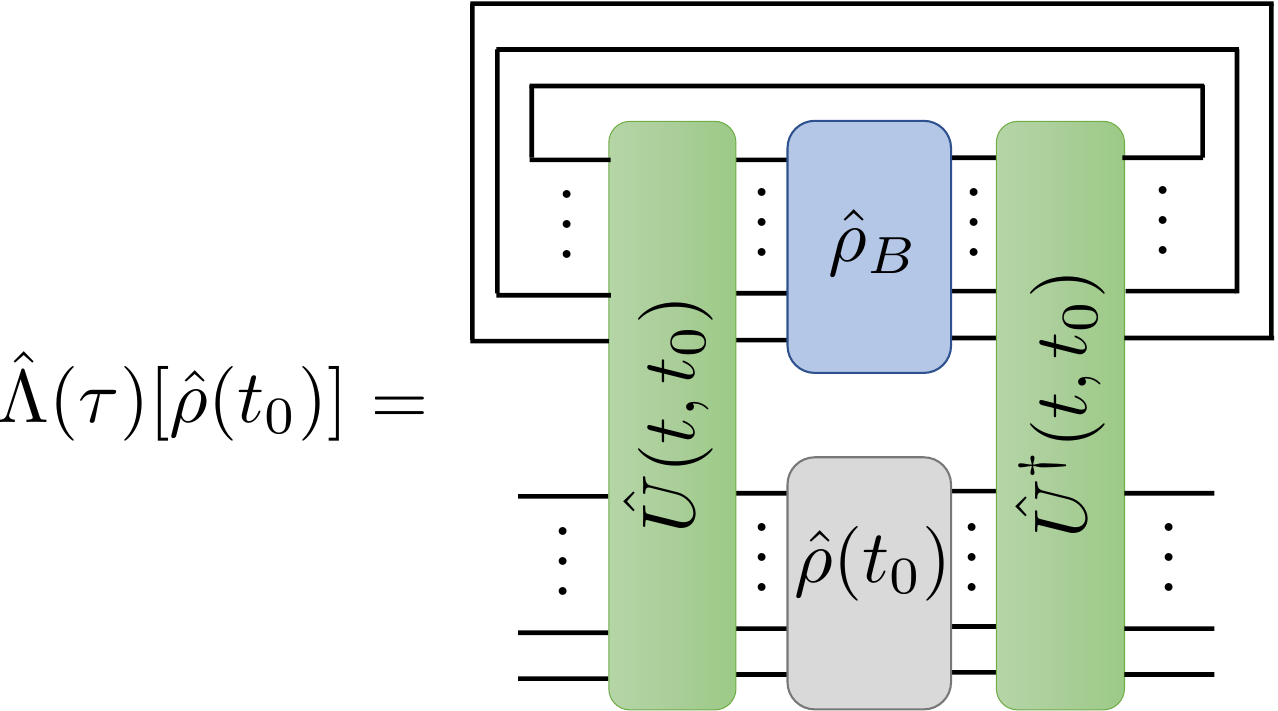}
  \caption{A schematic of the dynamical map $\hat{\Lambda}(\tau)$. The system $S$ and bath $B$ start in a product initial state $\hat{\rho}(t_0)\hat{\rho}_{B}$. The system and bath then undergo joint unitary evolution for a time duration $\tau$ after which the bath is traced out. This construction manifestly generates a CPTP map on the system's initial state~\cite{c8814692-9976-379a-9ad9-042fab94d853}.}
  \label{fig:dynamical_map}
\end{figure}

\section{Non-Markovian dynamics} \label{sec:nonmarkov}
The generic setup we analyse is a fermionic system with Hilbert space $\mathcal{H}_{S}$ connected to two non-interacting electronic reservoirs $\alpha=\{{\rm L},{\rm R}\}$, defined by equilibrium temperatures $T_{\alpha}=1/\beta_{\alpha}$ and chemical potentials $\mu_{\alpha}$ (taking $k_{B}=1=\hbar$ throughout) with Hilbert space $\mathcal{H}_{\alpha}$. The total Hamiltonian governing the dynamics of this system is given by 
\begin{equation} \label{eq:1}
\hat{H} = \hat{H}_S+\sum_{\alpha=L,R}\bigg(\hat{H}_{S\alpha}+\hat{H}_{\alpha}\bigg),
\end{equation}
where $\hat{H}_S$ is the system Hamiltonian, $\hat{H}_{\alpha}$ is the Hamiltonian of bath $\alpha$ and $\hat{H}_{S\alpha}$ describes the coupling between them. We restrict our analysis to Hamiltonians $\hat{H}$ that conserve fermion number $\hat{N}+\sum_{\alpha}\hat{N}_\alpha$ where $\hat{N}$ and $\hat{N}_{\alpha}$ are the total particle number operators for the system and bath $\alpha$ respectively. We study quench dynamics where the initial state of the full setup is taken to be of the product form $\hat{\rho}_{\textrm{tot}}(t_{0}) = \hat{\rho}(t_{0})\hat{\rho}_{B}$, where $\hat{\rho}_{B}$ is the composite initial thermal state of both baths, and $t_{0}$ is the initial time. The state of the system at a later time $t$ is then found by unitarily time evolving the system + bath and then tracing out the system bath as
\begin{align} 
\hat{\rho}(t) &= \hat{\Lambda}(t-t_{0})[\hat{\rho}(t_{0})], \nonumber \\
&= \textrm{Tr}_{B}\bigg(\hat{U}(t,t_0)\hat{\rho}(t_{0})\hat{\rho}_{B}\hat{U}^\dagger(t,t_0)\bigg), \label{eq:lambda}
\end{align}
where $U(t,t_0) = e^{-i\hat{H}(t-t_{0})}$, as depicted in Fig.~\ref{fig:dynamical_map}. From this construction $\hat{\Lambda}(t-t_{0})$ is manifestly a completely positive trace preserving (CPTP) map. A related exact description of open quantum system dynamics is given by the time convolutionless master equation \cite{nestmann2019timeconvolutionless,chruscinski2010non,
chruscinski2022dynamical}
\begin{align}
    \frac{\partial \hat{\rho}(\tau)}{\partial \tau}  = \hat{\mathcal{L}}(\tau)[\hat{\rho}(\tau)],
\end{align}
with time interval $\tau = t - t_0$ and
\begin{align}
\hat{\mathcal{L}}(\tau) = \frac{d}{d\tau}[\hat{\Lambda}(\tau)]\hat{\Lambda}^{-1}(\tau), \label{eq:propagator}
\end{align}
from which $\hat{\Lambda}(\tau) = \mathcal{T}e^{\int_{0}^{\tau}\hat{\mathcal{L}}(\tau')d\tau'}$, where $\mathcal{T}$ is the time-ordering operator. Note that $\hat{\mathcal{L}}(\tau)$ becomes undefined in the long-time limit as $\hat{\Lambda}(\tau)$ converges to a projector onto the steady state and thus has no inverse. 

We consider the situation where the system approaches a unique steady state in the long-time limit, i.e, $\lim_{\tau\to\infty}\hat{\Lambda}(\tau)[\hat{\rho}(0)]=\hat{\rho}(\infty)$ for any initial state $\hat{\rho}(0)$.  A difference between the above non-Markovian description and Markovian quantum dynamics is that, both $\hat{\Lambda}(\tau)$ and $\hat{\mathcal{L}}(\tau)$ are time-dependent, despite the global Hamiltonian $\hat{H}$ being time-independent. The instantaneous steady state, or the time-dependent fixed point, can be defined either as the eigenoperator of $\hat{\Lambda}(\tau)$ with eigenvalue $1$, or as the eigenoperator of $\hat{\mathcal{L}}(\tau)$ with eigenvalue $0$. Neither time-dependent fixed point may correspond to $\hat{\rho}(\infty)$ at short times, but both approach $\hat{\rho}(\infty)$ at long times. This leads to the existence of three dynamical timescales $\tau^{\Lambda}_{\rm m}$, $\tauL$ and $\tau_{\textrm{re}}$ given by
\begin{align} 
& ||\hat{\Lambda}(\tau)[\hat{\rho}(0)] - \hat{\rho}(\infty)|| <\epsilon,~~\forall~~\tau \geq\tau_{\textrm{re}}, \label{tau_re} \\
& ||\hat{\Lambda}(\tau)[\hat{\rho}(\infty)] - \hat{\rho}(\infty)|| <\epsilon,~~\forall~~\tau \geq\tau^{\Lambda}_{\rm m}, \label{tau_lambda} \\
& ||\hat{\mathcal{L}}(\tau)[\hat{\rho}(\infty)]||<\epsilon~~ \forall~~ \tau \geq\tauL, \label{tau_L} 
\end{align}
where $\epsilon$ is some arbitrarily small tolerance and $|| \hat{A}||$ is the norm of $\hat{A}$. Here $\tau_{\textrm{re}}$ defines the timescale of relaxation for an arbitrary initial state $\hat{\rho}(0)$. In contrast $\tauL$ defines the timescale over which a steady state can be well defined and $\tau^{\Lambda}_{\rm m}$ defines the timescale for which $\hat{\rho}(\infty)$ is dynamically invariant. These timescales are ordered as $\tauL \leq \tau^{\Lambda}_{\rm m} \leq \tau_{\textrm{re}}$, such that given access to $\hat{\Lambda}(\tau)$, and thus $\hat{\mathcal{L}}(\tau)$, for $\tau > \tau^{\mathcal{L}}_{\rm m}$, we can extract $\hat{\rho}(\infty)$ from dynamics over a memory timescale which can be orders of magnitude smaller than $\tau_{\rm re}$. Often it's numerically unfeasible to time evolve up to $\tau_{\rm re}$ due to the dynamical growth of entanglement, but by exploiting $\hat{\Lambda}(\tau)$ directly we can bypass this and accelerate the extraction of the steady state with significantly shorter time evolution. 

Crucially, we can also extract full transient dynamics with minimal computational cost in this dynamical map approach given the following assumption. If we assume the full propagator converges up to an error $O(\epsilon)$ on the timescale $\tau^{\mathcal{L}}_{\rm m}$, $\hat{\mathcal{L}}(\tau) = \hat{\mathcal{L}}_{\rm m}+O(\epsilon)$, $\forall \,\tau \geq \tau^{\mathcal{L}}_{\rm m }$ then we have the following factorised form for the dynamical map
\begin{equation} \label{eq:3}
\hat{\Lambda}(\tau) \approx e^{(\tau-\tauL)\hat{\mathcal{L}}_{\rm m}}\hat{\Lambda}(\tau^{\mathcal{L}}_{\rm m}) \quad \forall~~\tau\geq\tau^{\mathcal{L}}_{\rm m}.
\end{equation}
This decomposition represents a phenomenon called initial slippage which has been widely studied \cite{Bruch_2021,GEIGENMULLER198341,PhysRevA.28.3606,PhysRevA.32.2462,Gaspard1999SlippageOI} and is closely associated with the assumption of a finite memory time for the environment correlations \cite{PhysRevX.11.021041,Bruch_2021}. It also leads to an interesting anomalous thermal relaxation called the Non-Markovian Quantum Mpemba effect \cite{strachan2024nonmarkovian}. If we assume the dynamics is analytic and the system is finite-sized then we can use the key result in Ref.~\citen{Purkayastha_2021},
\begin{equation}
\hat{\rho}_{n\tau+t_{1}} = \underbrace{\hat{\Lambda}(\tau)[...[\hat{\Lambda}(\tau)}_{n~\textrm{times}}[\hat{\Lambda}(t_{1}-t_{0})\hat{\rho}(t_{0})]]]...], \nonumber 
\end{equation}
\begin{equation} \label{eq:4}
||\hat{\rho}(n\tau+t_{1})-\hat{\rho}_{n\tau+t_{1}}|| = \epsilon(\tau), ~\epsilon(\tau) \textrm{ decays with }\tau,
\end{equation}
where $t_{0}\leq{t}_{1}<\tau$ and $\tau^{\Lambda}_{\rm m}\ll\tau\leq \tau_{re}$. This can be thought of as an extension of repeated interaction models \cite{Purkayastha_2021} where each interaction of the system with the bath takes place over an extended timescale to capture non-Markovian effects rather than the standard instantaneous interaction repetition. 

In our approach, we can directly extract the dynamical map $\hat{\Lambda}(\tau)$ via the Choi-Jamiolkowski isomorphism and through Eq.~\eqref{eq:propagator} the propagator $\hat{\mathcal{L}}(\tau)$. Our focus in Sec.~\ref{sec:results} is on small systems $S$ comprising only a few fermionic modes, so given $\hat{\Lambda}(\tau)$ and $\hat{\mathcal{L}}(\tau)$ are superoperators acting only on system operators both can be constructed fully as a matrices. With full access to $\hat{\Lambda}(\tau)$ and $\hat{\mathcal{L}}(\tau)$ it is instructive to compute their spectral decomposition. Using the notation $(A|\circ := \textrm{Tr}_{S}A^{\dag}\circ$ and $|B) := B$ to denote the Hilbert-Schmidt scalar product $(A|B) = \textrm{Tr}_{S}A^{\dag}B$, their spectral decomposition follow as
\begin{align}
\hat{\Lambda}(\tau) &= \sum_{i=1}^{d^2}\Lambda_{i}(\tau)|g_{i}(\tau))(\bar{g}_{i}(\tau)|, \\
\hat{\mathcal{L}}(\tau) &= \sum_{i=1}^{d^2}\mathcal{L}_{i}(\tau)|h_{i}(\tau))(\bar{h}_{i}(\tau)|,
\end{align}
where $(\bar{g}_{i}|g_{j}) = \delta_{ij} = (\bar{h}_{i}|h_{j})$ and $d$ is the dimension of the system $S$ Hilbert space. Here $(\bar{g}_{i}|$ and $|g_{i})$ are distinct left and right eigenvectors of $\hat{\Lambda}(\tau)$ with the same eigenvalue $\Lambda_{i}$ and likewise is true of $(\bar{h}_{i}|$ and $|h_{i})$ for $\hat{\mathcal{L}}(\tau)$. As $\Lambda(\tau)$ is a CPTP map, we have $|\Lambda_{i}|\leq 1$ with a leading eigenvalue $\Lambda_{1}=1$ with $|g_{1}(\tau))=\hat{\rho}^{\Lambda}_{\rm FP}(\tau)$ corresponding to its time-dependent fixed point
\begin{align}
    \hat{\Lambda}(\tau)[\hat{\rho}^{\Lambda}_{\rm FP}(\tau)] &= \hat{\rho}^{\Lambda}_{\rm FP}(\tau).
\end{align}
We also have $\textrm{Re}(\mathcal{L}_{i}(\tau))\leq 0$ with a leading eigenvalue $\mathcal{L}_{1}(\tau)=0$ corresponding to $|h_{i}(\tau))=\hat{\rho}^{\mathcal{L}}_{\rm FP}(\tau)$  obeying
\begin{align}
    \hat{\mathcal{L}}(\tau)[\hat{\rho}^{\mathcal{L}}_{\rm FP}(\tau)] &= 0.
\end{align}
Here $\Lambda_{i}(\tau)$ can be interpreted as how much each mode $|g_{i}(\tau))$ has been damped by time $\tau$, with $\mathcal{L}_{i}(\tau)$ giving the corresponding instantaneous damping rate. Note that the modes themselves are also evolving, a behaviour not present in Markovian dynamics. In some cases where the dynamics is strongly non-Markovian, the decay of $|\Lambda_{i}(\tau)|$ is non monotonic and $\Lambda_{i}(\tau)=0$ can oscillate about zero before fully decaying away. This corresponds to a mode being suppressed and re-excited, caused by information backflow from the bath to the system. At these times $\hat{\mathcal{L}}(\tau)$ becomes ill-defined, but this poses no issues for times other than these isolated points~\cite{PhysRevA.104.062403,chruscinski2010non} and in fact $\hat{\mathcal{L}}(\tau)\hat{\Lambda}(\tau)$ is finite for all times~\cite{Nestmann:847387}. In these cases, the memory times must be chosen with care.

While formally, the stationary state can also be computed as the $\tau \to \infty$ limit of either of these fixed points, a key observation based on the PReB approach is that practically we only require that $\tau > \tau^\Lambda_{\rm m}$ or $\tau > \tau^\mathcal{L}_{\rm m}$, respectively. Equation~\eqref{eq:3} or \eqref{eq:4} can then be used to calculate the transient dynamics for $\tau>\tau^{\mathcal{L}}_{\rm m},\tau\gg\tau^{\Lambda}_{\rm m}$ to stationarity. We now proceed to outline the details of this approach.

\begin{figure}[t]
  \centering
  \includegraphics[width=0.35\textwidth]{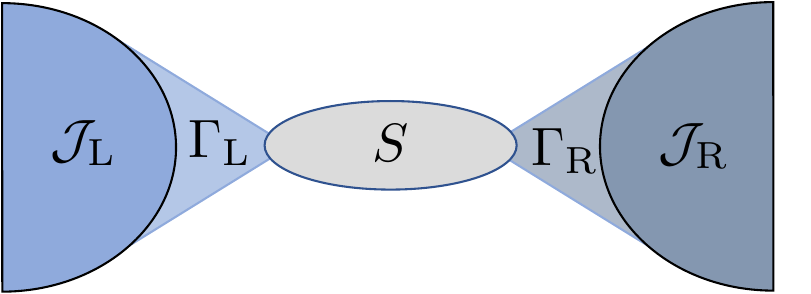}
  \caption{We consider a system $S$ comprising a small number of fermionic modes coupled to two infinite thermal Fermi baths $\alpha=\{{\rm L},{\rm R}\}$ with inverse temperatures $\beta_\alpha$, spectral density $\mathcal{J}_\alpha$ and coupling strength $\Gamma_\alpha$.}
  \label{fig:system_baths}
\end{figure}

\section{Modelling an infinite bath} \label{sec:baths}
We consider a setup as shown in Fig.~\ref{fig:system_baths} with a system coupled to two baths each being a continuum of non-interacting spinless fermionic modes~\footnote{The generalization to spin-1/2 fermions is straightforward, either by carrying an additional spin quantum number or doubling the number of spinless modes. The latter approach will be applied later in Sec.~\ref{sec:results}.} and together governed by
\begin{align}
    \hat{H}_{B}+\hat{H}_{SB} &= \sum_{\alpha=L,R}\Bigg(\int_{-D}^{D}\omega\hat{b}_{\alpha}^{\dag}(\omega)\hat{b}_{\alpha}(\omega){\rm d}\omega \\
    &\qquad + \int_{-D}^{D}\sqrt{\mathcal{J}_{\alpha}(\omega)}(\hat{b}_{\alpha}^{\dag}(\omega)\hat{Q}_{\alpha}+\hat{Q}^{\dag}_{\alpha}\hat{b}_{\alpha}(\omega)){\rm d}\omega\Bigg). \nonumber
\end{align}
Here for bath $\alpha$ we have its spectral density $\mathcal{J}_{\alpha}(\omega)$, the system operator coupling to it $\hat{Q}_{\alpha}$, and its canonical fermionic creation and annihilation operators $\hat{b}_{\alpha}^{\dag}(\omega),\hat{b}_{\alpha}(\omega)$ obeying $\{\hat{b}_{\alpha}^{\dag}(\omega),\hat{b}_{\alpha}(\omega')\}=\delta(\omega-\omega')$. Note that $\hat{Q}_{\alpha}$ and $\hat{b}_{\alpha}(\omega)$ also anticommute, $\{\hat{Q}_{\alpha},\hat{b}_{\alpha'}(\omega\} = 0$. We parameterise $\mathcal{J}_{\alpha}(\omega)$ via its total coupling strength 
\begin{align}
    \Gamma_{\alpha} = \frac{1}{2D}\int_{-D}^{D}2\pi\mathcal{J}_{\alpha}(\omega){\rm d}\omega,
\end{align}
where $D$ is its bandwidth of the baths, assumed to be identical. The initial state of the baths is a product state $\hat{\rho}_{B} = \prod_{\alpha}\hat{\rho}_{\alpha}$, where 
\begin{align}
    \hat{\rho}_{\alpha} = e^{-\beta_{\alpha}(\hat{H}_{\alpha}-\mu_{\alpha}\hat{N}_{\alpha})}/Z_{\alpha},
\end{align}
is a thermal state and $Z_{\alpha}$ is the partition function for bath $\alpha$.

\begin{figure}[t!]
    \centering
    \includegraphics[width=\columnwidth]{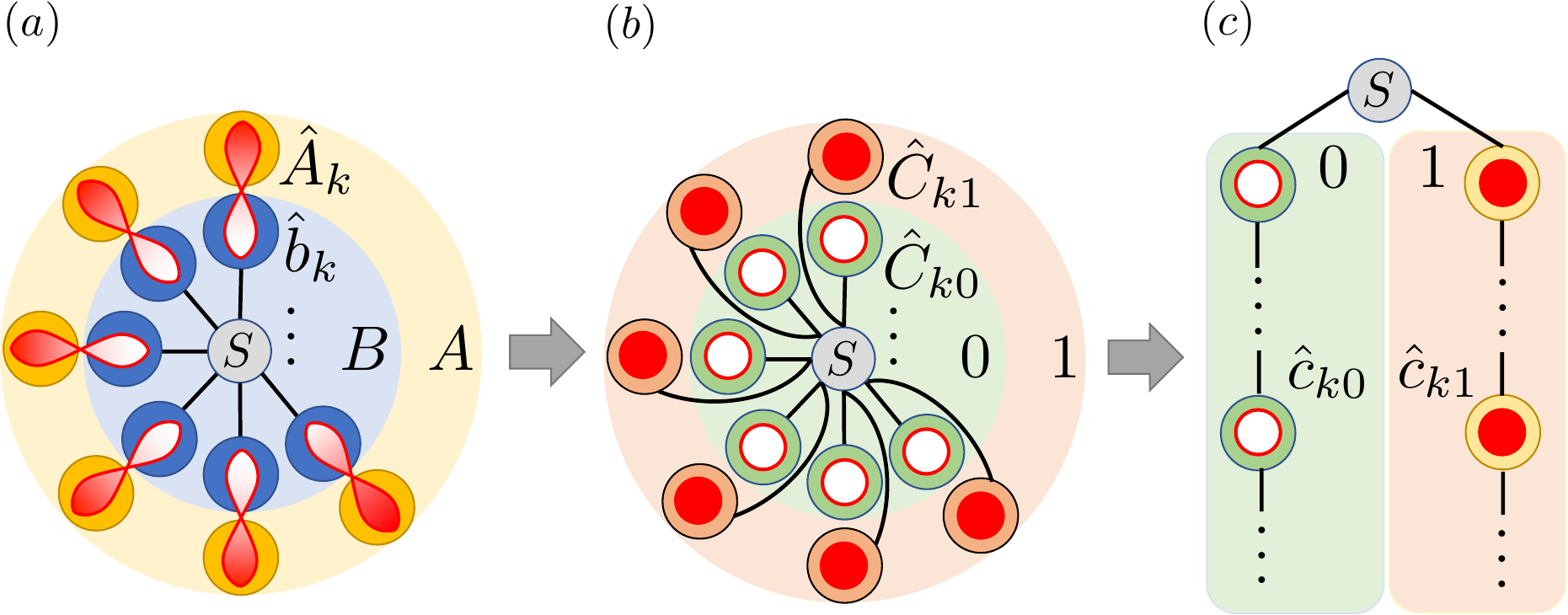}
    \caption{Illustration for a single bath, depicted as discrete modes, of the transformation into two chains. (a) Thermofield purification creates entangled states of each bath eigenmode and its corresponding auxiliary mode. Each bath eigenmode interacts with the system giving a star geometry, while the auxiliary modes are uncoupled. (b) Bath and auxiliary modes are mixed creating a set of empty ``0" and filled ``1" modes. All these modes couple to the system. (c) After tridiagonalizing the empty and filled modes separately we arrive at a two chain geometry better suited for tensor network calculations.}
    \label{fig:star_to_chain}
\end{figure}

In order to represent the equilibrium state of the bath as a pure state we use thermofield purification in which the finite temperature of any given bath is encoded in two different baths at zero temperature \cite{takahashi1996thermo}. For simplicity consider a single bath comprising a set of discrete modes governed by a non-interacting Hamiltonian $\hat{H}_{B}=\sum_{k}\epsilon_{k}\hat{b}^{\dag}_{k}\hat{b}_{k}$ at inverse temperature $\beta$ and chemical potential $\mu$, with each mode coupled to some system operator $\hat{Q}$ via a hybridisation Hamiltonian $\hat{H}_{SB} = \sum_{k}(v_{k}\hat{Q}^{\dag}\hat{b}_{k}+{v}^{*}_{k}\hat{b}^{\dag}_{k}\hat{Q})$. The thermal state follows as 
\begin{align}
\hat{\rho}_{\beta} &=\frac{e^{-\beta(\hat{H}_{B}-\mu\hat{N}_{B})}}{\textrm{Tr}(e^{-\beta(\hat{H}_{B}-\mu\hat{N}_{B})})} \nonumber \\
&= \prod_{k}\Big((1-f_{k})(\mathbbm{1} -\hat{b}^\dagger_k\hat{b}_k)+f_{k}\hat{b}^\dagger_k\hat{b}_k\Big),
\end{align}
where $f_{k} = (1+e^{\beta(\epsilon_{k}-\mu)})^{-1}$ is the Fermi factor for the $k^{\textrm{th}}$ mode. 
By introducing an auxiliary mode $\hat{A}_{k}$ for each bath mode $\hat{b}_{k}$, we can define the thermofield double state $\ket{\Omega_\beta}$ such that the thermal expectation value of an operator $\hat{O}$ acting on the system is given by $\bra{\Omega_\beta}\hat{O}\ket{\Omega_\beta} = \textrm{Tr}(\hat{\rho}_{\beta}\hat{O})$. This required state is then
\begin{equation}
\ket{\Omega_\beta} = \prod_k\bigg(\sqrt{1-f_{k}}\,\hat{A}^{\dag}_{k}+\sqrt{f_{k}}\,\hat{b}^{\dag}_{k}\bigg)\ket{\rm vac}, \label{eq:thermofield_state}
\end{equation}
for a bath with a star-geometry as depicted in Fig.~\ref{fig:star_to_chain}(a). Since purification is invariant to unitary transformations, $\ket{\Omega_{\beta}}$ is not unique. The form in Eq.~\eqref{eq:thermofield_state} has the advantage of being an eigenstate of the total particle number allowing its conservation to be exploited in numerical calculations.
By mixing the bath and auxiliary modes via a rotation 
\begin{equation}
\begin{pmatrix}
\hat{C}_{0,k} \\
\hat{C}_{1,k}
\end{pmatrix}
= 
\begin{pmatrix}
\textrm{cos}(\theta_{k}) & -\textrm{sin}(\theta_{k})  \\
\textrm{sin}(\theta_{k})  & \textrm{cos}(\theta_{k})  
\end{pmatrix}
\begin{pmatrix}
\hat{b}_{k} \\
\hat{A}_{k}
\end{pmatrix},
\end{equation}
where $\textrm{sin}(\theta_{k}) = \sqrt{f_{k}}$, we map $\ket{\Omega_{\beta}}$ to a Fock state
\begin{align}
    \ket{\Omega_{\beta}} = \prod_{k}\hat{C}^\dagger_{1,k}\ket{\textrm{vac}},
\end{align}
in which the ``0" modes $\hat{C}_{0,k}$ are completely empty while the ``1" modes $\hat{C}_{1,k}$ are completely filled. As the physical and auxiliary modes are decoupled in the Hamiltonian, we are free to choose an arbitrary Hamiltonian for the auxiliary modes themselves. A convenient choice is to set $\hat{H}_{A} = \sum_{k}\epsilon_{k}\hat{A}^{\dag}_{k}\hat{A}_{k}$ which leads to a mixed mode Hamiltonian
\begin{equation}
\hat{H}_{A}+\hat{H}_{B} = \sum_{k}\epsilon_{k}(\hat{C}^{\dag}_{0,k}\hat{C}_{0,k}+\hat{C}^{\dag}_{1,k}\hat{C}_{1,k}).
\end{equation}
Due to the mixing all modes now couple to the system as 
\begin{align}
\hat{H}_{SB} &= \sum_{k}\sqrt{f_k}\Big(v_{k}\hat{Q}^{\dag}\hat{C}_{1,k}+{v}^{*}_{k}\hat{C}^{\dag}_{1,k}\hat{Q}\Big) \nonumber\\
& \qquad - \sum_{k}\sqrt{1-f_k}\Big(v_{k}\hat{Q}^{\dag}\hat{C}_{0,k}+{v}^{*}_{k}\hat{C}^{\dag}_{0,k}\hat{Q}\Big),
\end{align}
giving a modified star geometry bath shown in Fig.~\ref{fig:star_to_chain}(b) and in which the thermal properties of the initial state have now been encoded.

\begin{figure}[t!]
  \centering
  \includegraphics[width=0.5\textwidth]{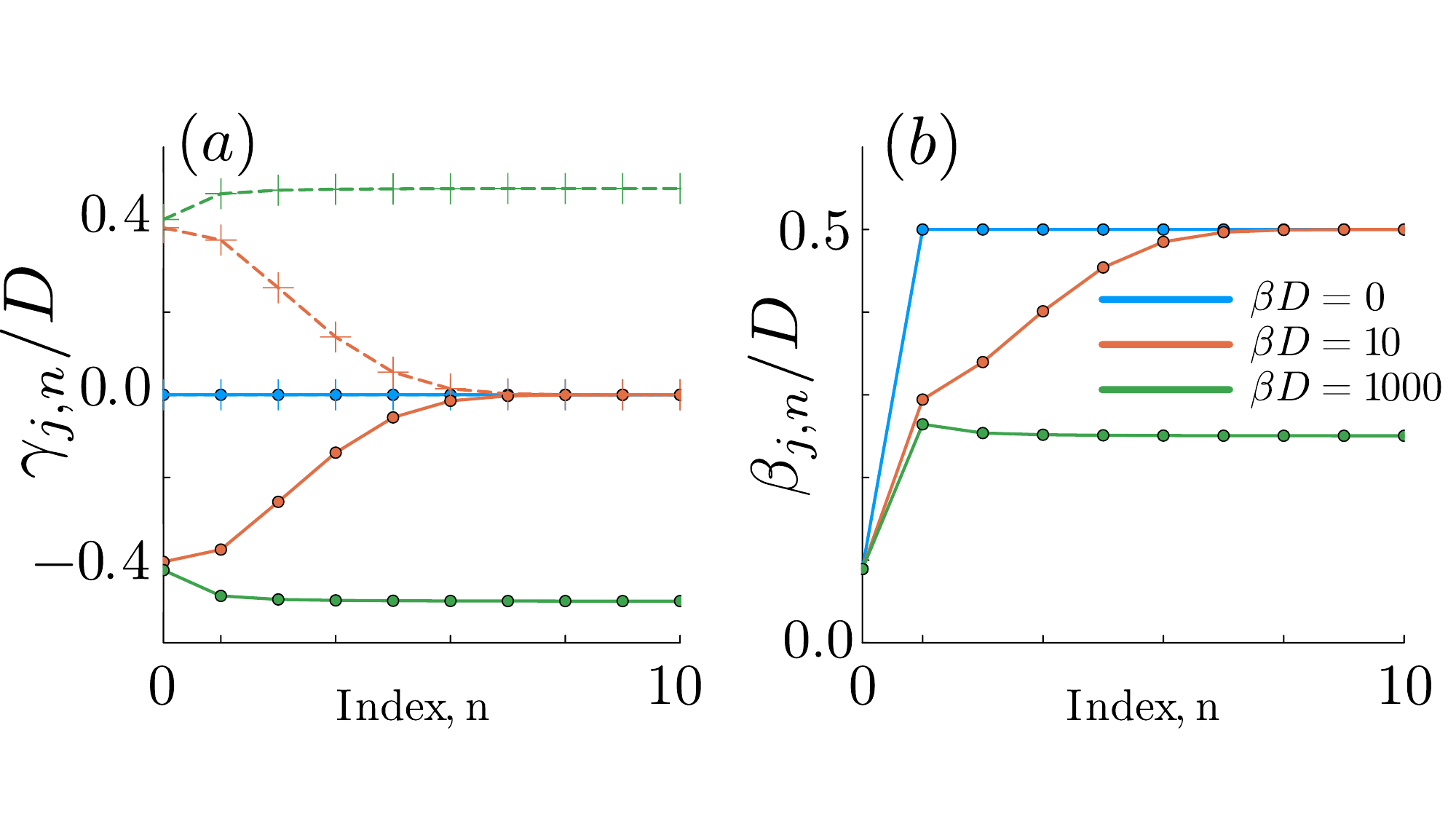}
  \caption{Convergence of chain coefficients for the orthogonal polynomial mapping for a single bath with $\mu=0,\Gamma=0.05D$ and a semi-elliptical spectral function defined as Eq.~(\ref{eq:elliptical spectral function}). (a) Energies and (b) couplings along the empty ``0" (solid) and fully occupied ``1" (dashed) chains. Due to the symmetric spectral density $\mathcal{J}(\omega)=\mathcal{J}(-\omega)$ we have $\beta_{0,n}=\beta_{1,n}$ and $\gamma_{0,n}=-\gamma_{1,n}$.}
  \label{fig: convergence of chain properties}
\end{figure}

Moving back to the continuum with two baths $\alpha$ we have $v_{k}\to \sqrt{\mathcal{J}_{\alpha}(\omega)}$, $f_{k}\to f_{\alpha}(\omega)$, $\hat{C}_{j,k}\to \hat{C}_{\alpha j}(\omega)$, the equivalent thermofield mixed Hamiltonian is 
\begin{align}
\hat{H} &= \hat{H}_{S}+\sum_{\alpha j}\int_{-D}^{D}{\rm d}\omega\,\omega\hat{C}_{\alpha j}^{\dag}(\omega)\hat{C}_{\alpha j}(\omega) \nonumber \\
&\qquad+ \sqrt{\mathcal{J}_{\alpha j}(\omega)}\big[\hat{C}_{\alpha j}^{\dag}(\omega)\hat{Q}_{\alpha} + \hat{Q}_{\alpha}^{\dag}\hat{C}_{\alpha j}(\omega) \big], 
\end{align}
where the empty and filled spectral densities are
\begin{align}
    \mathcal{J}_{\alpha 0}(\omega) &= (1-f_{\alpha}(\omega))\mathcal{J}_{\alpha}(\omega), \\
    \mathcal{J}_{\alpha 1}(\omega) &= f_{\alpha}(\omega)\mathcal{J}_{\alpha}(\omega),
\end{align}
and $\ket{\Omega_{\beta}} = \prod_{\alpha}\prod_{-D}^{D}\hat{C}^{\dag}_{\alpha 1}(\omega)\ket{\textrm{vac}}$ is a continuous Fock state. To render this problem practicable for numerical calculations the continuous baths are discretized and truncated into a finite number of modes. In some applications this may be accomplished by assuming some linear or logarithmic binning in the energy axis of the baths~\cite{RevModPhys.80.395}. Here our focus is on capturing the time-evolution as accurately as possible up to some prescribed time. This is accomplished by performing a continuous mode tridiagonalization of the two zero temperature star geometry baths into two one-dimensional tight-binding chains as shown in Fig.~\ref{fig:star_to_chain}(c). In doing this we define a new set of fermionic operators $\hat{c}_{\alpha j,n}$ as~\cite{Chin2010}
\begin{equation}
\hat{C}_{\alpha j}(\omega) = \sum_{n}U_{\alpha j,n}(\omega)\hat{c}_{\alpha j,n},
\end{equation}
where
\begin{align}
    U_{\alpha j,n}(\omega) &= \frac{\sqrt{\mathcal{J}_{\alpha j}(\omega)}\pi_{\alpha j,n}(\omega)}{\rho_{\alpha j, n}},
\end{align}
for $j=0,1$ and $\pi_{\alpha j,n}(\omega)$ are monic orthogonal polynomials that obey 
\begin{align}
    \int_{-D}^{D}{\rm d}\omega \,\mathcal{J}_{\alpha j}(\omega)\pi_{\alpha j,n}(\omega)\pi_{\alpha j,m}(\omega) &=\rho_{\alpha j,n}^{2}\delta_{nm},
\end{align}
with 
\begin{align}
    \rho_{\alpha j,n}^{2} &= \int_{-D}^{D}{\rm d}\omega\,\mathcal{J}_{\alpha j}(\omega)\pi_{\alpha j,n}^{2}(\omega).
\end{align}
Using a finite cutoff of $M$ modes for the mapping, we arrive at 
\begin{align}
\hat{H} &= \hat{H}_{S} + \sum_{\alpha j}\bigg[\rho_{\alpha j,0}(\hat{Q}^{\dag}_{\alpha}\hat{c}_{\alpha j,0}+\hat{c}_{\alpha j,0}^{\dag}\hat{Q}_{\alpha})  \nonumber \\ 
&\qquad\qquad\quad  +
\sum_{n=0}^{M}\gamma_{\alpha j,n}\hat{c}^{\dag}_{\alpha j,n}\hat{c}_{\alpha j,n}  \nonumber \\
&\qquad\qquad\quad + (\sqrt{\beta_{\alpha j,n+1}}\hat{c}^{\dag}_{\alpha j,n+1}\hat{c}_{\alpha j,n}+ {\rm H.c.})\bigg], \label{eq:chain_hamiltonian}
\end{align}
where the coefficients $\gamma_{\alpha j,n}$ and $\beta_{\alpha j,n}$ are defined through the recurrence relation
\begin{align}
    \pi_{\alpha j,n+1}(k) &= (k-\gamma_{\alpha j,n})\pi_{\alpha j,n}(k)-\beta_{\alpha j,n}\pi_{\alpha j,n-1}(k), \nonumber
\end{align}
with $\pi_{\alpha j,-1}(k) = 0$. These chain parameters are generated using the \texttt{ORTHPOL} package~\cite{Gautschi2005}. Generically, they are found to quickly converge to constants $\gamma_{\alpha j,n}\to\gamma_{\alpha j}$, $\beta_{\alpha j,n}\to\beta_{\alpha j}$. For a weighting function $\mathcal{J}_{\alpha j}(\omega)$ with support $[a,b]$ it can be shown~\cite{Rosenbach_2016} that these coefficients converge as $\gamma_{\alpha j,n} \to (a+b)/2$ and $\beta_{\alpha j,n} \to (b-a)^{2}/16$ for $n\to \infty$. In Fig.~\ref{fig: convergence of chain properties} the chain coefficients $\beta_{j,n},\gamma_{j,n}$ are shown for a single bath with with a semi-elliptical spectral function defined as Eq.~(\ref{eq:elliptical spectral function}) and $\mu=0$ for various temperatures. At zero temperature, $\mathcal{J}_{j}(\omega)$ has support $[0,D],[-D,0]$ for $j=0,1$ respectively and for $T>0$ both have support $[-D,D]$. For the case shown we see that the chain coefficients converge after only a few sites to the asymptotic values expected from theory. By employing the Lieb-Robinson bound \cite{Woods_2016} sites further than $\sim \tau\beta_{\alpha j}$ in the $\alpha j$ chain have a negligible effect on the system dynamics up to time $\tau$, giving a well-defined measure of the length of bath chains needed to accurately capture the dynamics. In this sense, the discretization of the baths generated by orthogonal polynomials up to some length $M$ is an exact description of the infinite bath up to this finite time. 

\section{Extracting the dynamical map} \label{sec:dynamical}
Our key methodological proposal here is to compute and exploit directly the dynamical map $\hat{\Lambda}(\tau)$ for open systems. To accomplish this we use the Choi-Jamiolkowski isomorphism which provides an equivalence between quantum states and superoperators~\cite{CHOI1975285,JAMIOLKOWSKI1972275}. 

\begin{figure}[t]
  \centering
  \includegraphics[width=0.45\textwidth]{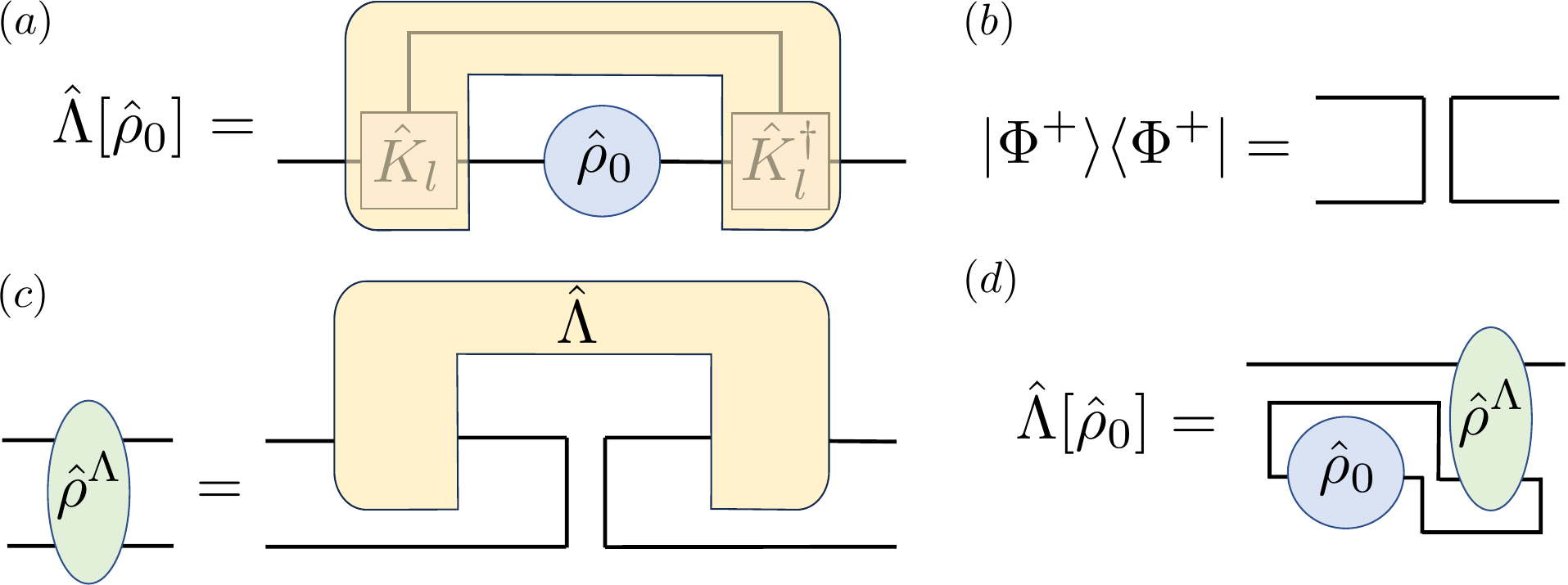}
  \caption{Tensor network type diagrams of the Choi-Jamiolkowski isomorphism. (a) The application of a CPTP map $\hat{\Lambda}[\hat{\rho}_0]$ can be viewed as cap shaped tensor encompassing $\hat{\rho}$. Internally the cap tensor could be decomposed into a conjugation by Kraus operators $\hat{K}_l$ as shown. (b) Owing to its perfect correlations the maximally entangled state $\ket{\Phi^+}$ from Eq.~\eqref{eq:maximally_entangled} is a bent line. (c) The Choi-Jamiolkowski isomorphism follows from inputting $\ket{\Phi^+}$ into $\hat{\Lambda}$, which diagrammatically corresponds to bending the input legs of $\hat{\Lambda}$ outwards to form $\hat{\rho}^\Lambda$ as in Eq.~\eqref{eq:isomorphism_state}. (d) The application of the map $\hat{\Lambda}[\hat{\rho}_0]$ is extracted from $\hat{\rho}^\Lambda$ by contracting the appropriate legs consistent with (a), and inserting a factor of the dimension $d$ to get the graphical version of Eq.~\eqref{eq:choi_map}.}
  \label{fig:isomorphism}
\end{figure}

\subsection{Choi-Jamiolkowski isomorphism}
Mirroring the thermofield construction for the bath the isomorphism starts by adding auxilliary replica degrees of freedom $A_S$ for the system $S$ as well. Consider a $d$-level system each with basis states $\ket{s}$ for which the application of any CPTP map $\hat{\Lambda}$ to an initial state $\hat{\rho}_{0}$ can be drawn as in Fig.~\ref{fig:isomorphism}(a). The trick of the isomorphism is to initialize the system and its auxiliary replica in a perfectly correlated maximally entangled state
\begin{align}
   \ket{\Phi^{+}} = \frac{1}{\sqrt{d}}\sum_{s=1}^{d}\ket{s}_{S}\otimes\ket{s}_{A_S}, \label{eq:maximally_entangled}
\end{align}
who's diagrammatic form is shown in Fig.~\ref{fig:isomorphism}(b). Since $\hat{\Lambda}$ is completely positive applying it to the system alone is guaranteed to generate a valid density operator in the space of system + replica operators~\footnote{Conversely, for any nonnegative operator we can also to a corresponding quantum map.} given by
\begin{align}
    \hat{\rho}^{\Lambda} = (\hat{\Lambda}\otimes\mathds{1})\{\ket{\Phi^{+}}\bra{\Phi^{+}}\}, \label{eq:isomorphism_state}
\end{align}
as depicted in Fig.~\ref{fig:isomorphism}(c). Armed with $\hat{\rho}^{\Lambda}$ the action of the map $\hat{\Lambda}$ then follows
\begin{equation}
    \hat{\Lambda}[\hat{\rho_{0}}] = d\;\textrm{Tr}_{A_S}(\mathbb{1}\otimes\hat{\rho}_{0}^{\textrm{T}}\hat{\rho}^{\Lambda}), \label{eq:choi_map}
\end{equation}
as summarized in Fig.~\ref{fig:isomorphism}(d). The isomorphism straightforwardly extends to maps applied to $N$ system sites by using the state
\begin{align}
   \ket{\Phi^{+}_N} = \frac{1}{\sqrt{d^N}}\otimes_{j=1}^N\left(\sum_{s_j=1}^{d}\ket{s_j}_{S}\otimes\ket{s_j}_{A_S}\right), 
\end{align}
where each site is maximally entangled with its own auxiliary replica. The application of the isomorphism to many-body fermionic systems also follows analogously once some subtle details are taken care of as we now describe.

\subsection{Application to fermionic systems} \label{sec:fermion_systems}
\begin{figure}[t]
  \centering
  \includegraphics[width=0.35\textwidth]{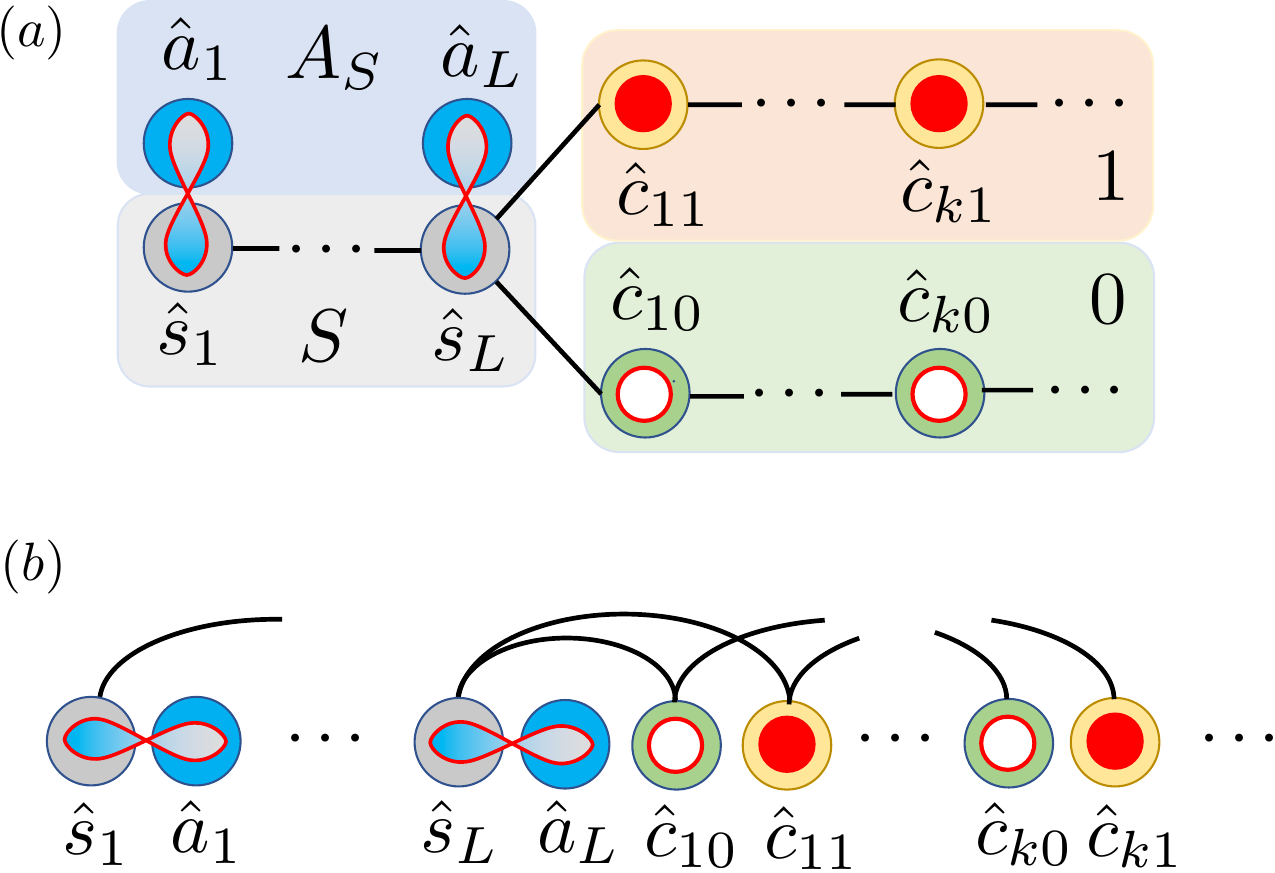}
  \caption{(a) For the case of a single bath our setup comprises the system $S$, it auxiliary replica $A_S$ and two thermofield chains ``0" and ``1" which are empty and filled, respectively. (b) The assumed fermionic mode ordering used in this work. The system and its replicas are interleaved, but separated from the thermofield chains which are also interleaved together. In this ordering local interactions between modes now become 3-local (next-nearest-neighbour) and 4-local interactions along the one-dimensional ordering of modes.}
  \label{fig:mode_ordering}
\end{figure}

Take the system $S$ as being described by $L$ spinless fermionic modes $\hat{s}_j$. We then additionally introduce corresponding auxiliary modes $\hat{a}_j$ for each of them and anticipate that these pairs of modes will be initialized in an entangled state. For a single bath the setup has the form shown in Fig.~\ref{fig:mode_ordering}(a). To construct fermionic Fock states for the setup a specific one-dimensional mode ordering is needed. We will consider two such orderings. The first is {\em separated} ordering  
\begin{align}
\{\hat{d}_{i}\}_{\rm sep} &= \{\hat{s}_{1},\cdots,\hat{s}_{L},\hat{a}_{1},\cdots,\hat{a}_{L}, \nonumber \\
& \qquad\qquad\qquad\qquad \hat{c}_{0,1},,\cdots,\hat{c}_{0,M},\hat{c}_{1,1},\cdots,\hat{c}_{1,M}\}, \nonumber
\end{align}
where system, auxiliary replica and each set of thermofield bath modes are grouped together. Fock states of the setup then follow as
\begin{align}
\ket{\boldsymbol{n}^S,\boldsymbol{n}^{A_S},\boldsymbol{n}^{0},\boldsymbol{n}^1} &= (\hat{s}_1^\dagger)^{n^S_1}\cdots (\hat{s}_L^\dagger)^{n^S_L} (\hat{a}_1^\dagger)^{n^{A_S}_1}\cdots (\hat{a}_L^\dagger)^{n^{A_S}_L} \nonumber \\
&\qquad\quad\times (\hat{c}_{0,1}^\dagger)^{n^{0}_1}\cdots (\hat{c}_{0,M}^\dagger)^{n^{0}_M}\nonumber \\
&\qquad\qquad\quad \times(\hat{c}_{1,1}^\dagger)^{n^{1}_{1}}\cdots(\hat{c}_{1,M}^\dagger)^{n^{1}_{M}}\ket{\rm vac}, \nonumber
\end{align}
with $\boldsymbol{n}^S = \{0,1\}^L$ being the occupation numbers of the system modes, while $\boldsymbol{n}^{A_S}$ and $\boldsymbol{n}^{0},\boldsymbol{n}^{1}$ are likewise for the replica modes and the ``0"/``1" thermofield chains.

One issue for the application of the Choi-Jamiolkowski to fermions is that it formally requires a partial trace over fermionic modes. Since fermionic Fock states are constructed from mode operators with a nonlocal canonical anticommutator algebra they lack an underlying tensor product structure associated with the conventional partial trace~\cite{Amosov_2016,Cirio_2022}. However, it can be shown that if the modes to be retained are a contigiuous subset of the assumed mode ordering, and the overall state is number parity symmetric, then the conventional partial trace coincides with the fermionic mode-reduced states, as shown explicitly in Ref.~\citen{Amosov_2016}. The separated mode ordering accomplishes this for the thermofield bath modes, which are traced out in the definition of $\hat{\Lambda}(\tau)$ in Eq.~\eqref{eq:lambda}, and for the replica modes of the system which are traced out in Eq.~\eqref{eq:choi_map} when reconstructing $\hat{\Lambda}(\tau)$ from $\hat{\rho}^\Lambda(\tau)$. 

With this ordering in place the Choi-Jamiolkowski isomorphism proceeds by initializing the setup in the state
\begin{align}
    \ket{\Psi_{\rm CJ}} = \sum_{\boldsymbol{n}\in\{0,1\}^L} \frac{1}{\sqrt{2^L}}\ket{\boldsymbol{n},\boldsymbol{n},0,\dots,0,1,\dots,1},
\end{align}
where the occupation of the system and replica modes are perfectly correlated while the thermofield chains are empty and filled. To extract $\hat{\Lambda}(\tau)$ this initial state is time-evolved as $\ket{\Psi_{\rm CJ}(\tau)} = \exp(-i\hat{H}\tau)\ket{\Psi_{\rm CJ}}$ from which we obtain $\hat{\rho}^{\Lambda}(\tau) = \textrm{Tr}_{01}(\ket{\Psi_{\rm CJ}(\tau)}\bra{\Psi_{\rm CJ}(\tau)})$. 

While the state $\ket{\Psi_{\rm CJ}}$ is number-parity symmetric it does not possess a fixed total particle number. When $\hat{H}$ itself is number conserving, as it is here, it is numerically advantageous to fully exploit this symmetry. To do this we use a modified initial state with the system and replica modes having an anti-correlated occupation
\begin{align} \label{eq:anti_correlated_state}
    \ket{\Psi_{\rm AC}} = \sum_{\boldsymbol{n}\in\{0,1\}^L} \frac{1}{\sqrt{2^L}}\ket{\boldsymbol{n},\bar{\boldsymbol{n}},0,\dots,0,1,\dots,1},
\end{align}
where $\bar{n}_j = n_j \oplus 1$ denotes the occupation flipping of a mode. This state over the system, replica and thermofield modes is precisely half-filled. We are permitted to use $\ket{\Psi_{\rm AC}}$ in place of $\ket{\Psi_{\rm CJ}}$ and compute $\hat{\rho}^{\rm AC}(\tau) = \textrm{Tr}_{01}(\ket{\Psi_{\rm AC}(\tau)}\bra{\Psi_{\rm AC}(\tau)})$ since there exists a unitary transformation $\hat{P}\ket{\Psi_{\rm AC}} = \ket{\Psi_{\rm CJ}}$ which acts exclusively on the replica modes such that $[\hat{P},\hat{H}]=0$, as outlined in Appendix~\ref{appendix: ancilla unitaries}. Consequently, 
\begin{align*}
    \hat{\rho}^{\Lambda}(\tau) &= \textrm{Tr}_{01}(e^{-i\hat{H}\tau}\ket{\Psi_{\rm CJ}}\bra{\Psi_{\rm CJ}}e^{i\hat{H}\tau}), \\
   &= \textrm{Tr}_{01}(e^{-i\hat{H}\tau}\hat{P}\ket{\Psi_{\rm AC}}\bra{\Psi_{\rm AC}}\hat{P}^\dagger e^{i\hat{H}\tau}), \\
&= \hat{P}\hat{\rho}^{\rm AC}(\tau)\hat{P}^{\dag}, 
\end{align*}
and hence $\hat{P}$ can be applied to the reduced state of the system and its replica after the time-evolution with the bath has been performed. 

The overall setup generalises straightforwardly to two baths $\alpha = \{{\rm L},{\rm R}\}$ by introducing the two thermofield chains for the left bath at the start as 
\begin{align}
\{\hat{d}_{i}\}_{\rm sep} &= \{\hat{c}_{{\rm L}0,1},,\cdots,\hat{c}_{{\rm L}0,M},\hat{c}_{{\rm L}1,1},\cdots,\hat{c}_{{\rm L}1,M}, \\
&\qquad\qquad\hat{s}_{1},\cdots,\hat{s}_{L},\hat{a}_{1},\cdots,\hat{a}_{L}, \nonumber \\
& \qquad\qquad\qquad \hat{c}_{{\rm R}0,1},\cdots,\hat{c}_{{\rm R}0,M},\hat{c}_{{\rm R}1,1},\cdots,\hat{c}_{{\rm R}1,M}\}, \nonumber
\end{align}
Otherwise the calculation proceeds identically.

\subsection{Non-interacting fermions} \label{sec:noninteracting}
If $\hat{H}_{S}$ is quadratic in mode operators and particle number conserving, then the exact dynamics can be obtained via unitary evolution of the single-particle correlation matrix $\boldsymbol{C}_{ij}(\tau) = \textrm{Tr}(\hat{\rho}(\tau)\hat{d}^{\dag}_{j}\hat{d}_{i})$ over all the system, bath and auxiliary modes $\{\hat{d}_i\}$. Given $\hat{H} = \sum_{ij}\boldsymbol{h}_{ij}\hat{d}^{\dag}_{i}\hat{d}_{j}$ the time-evolved correlation matrix follows from
\begin{equation} \label{eq:20}
\boldsymbol{C}(\tau) = e^{i\boldsymbol{h}\tau}\boldsymbol{C}(t_{0})e^{-i\boldsymbol{h}\tau},
\end{equation}
as shown in detail in Appendix~\ref{appendix: corr propagation}. The reduced density matrix of the system and its auxiliary modes is directly obtained from $\boldsymbol{C}(\tau)$ via \cite{cheong2004many}
\begin{align}
\hat{\rho}^{\rm AC}(\tau) &= \mathrm{det}(\mathbb{1}-\boldsymbol{C}^{\rm T}(\tau)) \label{eq:21} \\
& \times \mathrm{exp}\bigg\{\sum_{ij \in SA_S}[\mathrm{log}(\boldsymbol{C}^{\rm T}(\tau))(\mathbb{1}-\boldsymbol{C}^{\rm T}(\tau))^{-1}]_{ij}\hat{d}^{\dag}_{i}\hat{d}_{j}\bigg\}, \nonumber
\end{align}
where $SA_S$ denotes the $2L$ modes spanning the system and its replica modes. This equation holds provided $\hat{\rho}^{\rm AC}(\tau)$ is block diagonal in the number basis, as is the case here for the particle number symmetric initial state $\ket{\Psi^{\rm AC}}$. After extracting $\hat{\rho}^{\Lambda}(\tau)$ a highly non-trivial verification of its correctness is achieved by comparing to the prediction of Landauer-B\"{u}ttiker theory, as presented in Sec.~\ref{sec:results}.

\subsection{Matrix product state time evolution}
To handle interacting fermions we employ matrix product state (MPS) techniques. Given a many-body system of $N$ qubits each described by Pauli-$Z$ eigenstates $\ket{\sigma}$ with $\sigma = \pm 1$, a pure state in its tensor product Hilbert space $(\mathbbm{C}^2)^{\otimes N}$ is represented by an MPS as~\cite{PhysRevLett.91.147902,10.5555/2011832.2011833}
\begin{align}
\ket{\psi} &= \sum_{\sigma_{1},\dots,\sigma_{N}}A^{\sigma_{1}}_{1}A^{\sigma_{2}}_{2}\cdots A^{\sigma_{N-1}}_{N-1}A^{\sigma_{N}}_{N}\ket{\sigma_{1},\dots,\sigma_{N}},
\end{align}
where $A_{j}^{\sigma_{j}}$ is a tensor formed from $\chi_{j-1}\times\chi_{j}$ matrix (with $\chi_{0}=\chi_{N}=1$ fixed) indexed by the state $\sigma_j$ of the $j$th qubit. A crucial feature of MPS is that the bond dimension $\chi_j$ can be directly linked to the entanglement between the two halves of the system it connects~\cite{ORUS2014117}. The one-dimensional connectivity of MPS makes them naturally suited to describing systems with Hamiltonians sharing this geometry where an area law of entanglement is expected to severely limit the entanglement in its ground state and low-lying excitations~\cite{Hastings_2007}. 

The chain-like geometry of the open systems we arrived at in Eq.~\eqref{eq:chain_hamiltonian} and Fig.~\ref{fig:mode_ordering}(a) thus appears to be well suited to MPS. However, the separated mode ordering introduced in Sec.~\ref{sec:fermion_systems} is suboptimal for MPS, so we instead adopt {\em interleaved} ordering. For a two-bath setup this is given by 
\begin{align}
\{\hat{d}_{i}\}_{\rm int} &= \{\hat{c}_{{\rm L}1,M},\hat{c}_{{\rm L}0,M},\cdots,\hat{c}_{{\rm L}1,1},\hat{c}_{{\rm L}0,1}, \label{eq:ordering} \\
&~\quad \hat{s}_{1},\hat{a}_{1},\cdots,\hat{s}_{L},\hat{a}_{L},\hat{c}_{{\rm R}0,1},\hat{c}_{{\rm R}1,1},\cdots,\hat{c}_{{\rm R}0,M},\hat{c}_{{\rm R}1,M}\}. \nonumber
\end{align}
In this ordering the thermofield chains for each baths are interleaved, as are the system and replica modes. This mode ordering has two key advantages. First, it minimizes the length of interactions by avoiding terms coupling modes spanning across separated chains. This results in the nearest-neighbour interactions in the chains becoming next-nearest-neighbour interactions in the MPS, as shown in Fig.~\ref{fig:mode_ordering}(b). Locality is especially important since spinless fermionic modes are described in a systems of qubits using the Jordan-Wigner transformation~\cite{10.1119/1.1482064,Nielsen2005TheFC}. Consequently long-ranged fermionic couplings become long-ranged multi-body operators in the equivalent qubit Hamiltonian that are potentially deleterious due to dynamical entanglement growth. Second, interleaved ordering minimizes the MPS bond dimension required to capture the initial system + replica mode entangled state in Eq.~\eqref{eq:anti_correlated_state} by locating the entangled mode pairs adjacent to one another. In contrast, the separated ordering results in ``rainbow"-like state with a bond dimension $\chi_j \sim 2^L$.

The MPS computation of the time-evolution of $\ket{\Psi_{\rm AC}}$ in interleaved mode ordering proceeds using the two-site time-dependent variational principle (2TDVP) algorithm~\cite{Yang_2020,Fishman_2022,Haegeman_2016} exploiting particle number conservation. While the effective qubit Hamiltonian is short-ranged it does have terms beyond nearest-neighbour so it is essential to use a global subspace expansion to reduce the projection error~\cite{PhysRevB.102.094315}. Our calculations were all performed using the \texttt{ITensor} package~\cite{Fishman_2022}. Since the interleaved ordering still separates the bath modes from the system and its replica modes the computation of $\hat{\rho}^{\rm AC}(\tau)$ is unchanged. The final extraction of $\hat{\Lambda}(\tau)$ though requires that the system and replica modes are swapped back into separated ordering as detailed in Appendix~\ref{appendix: ancilla unitaries}. This step is potentially inefficient for large $L$ and avoiding it is the subject of ongoing work~\cite{Future_work}. For the small systems considered here this presents no issues. 

\begin{figure}[t]
  \centering
  \includegraphics[width=0.48\textwidth]{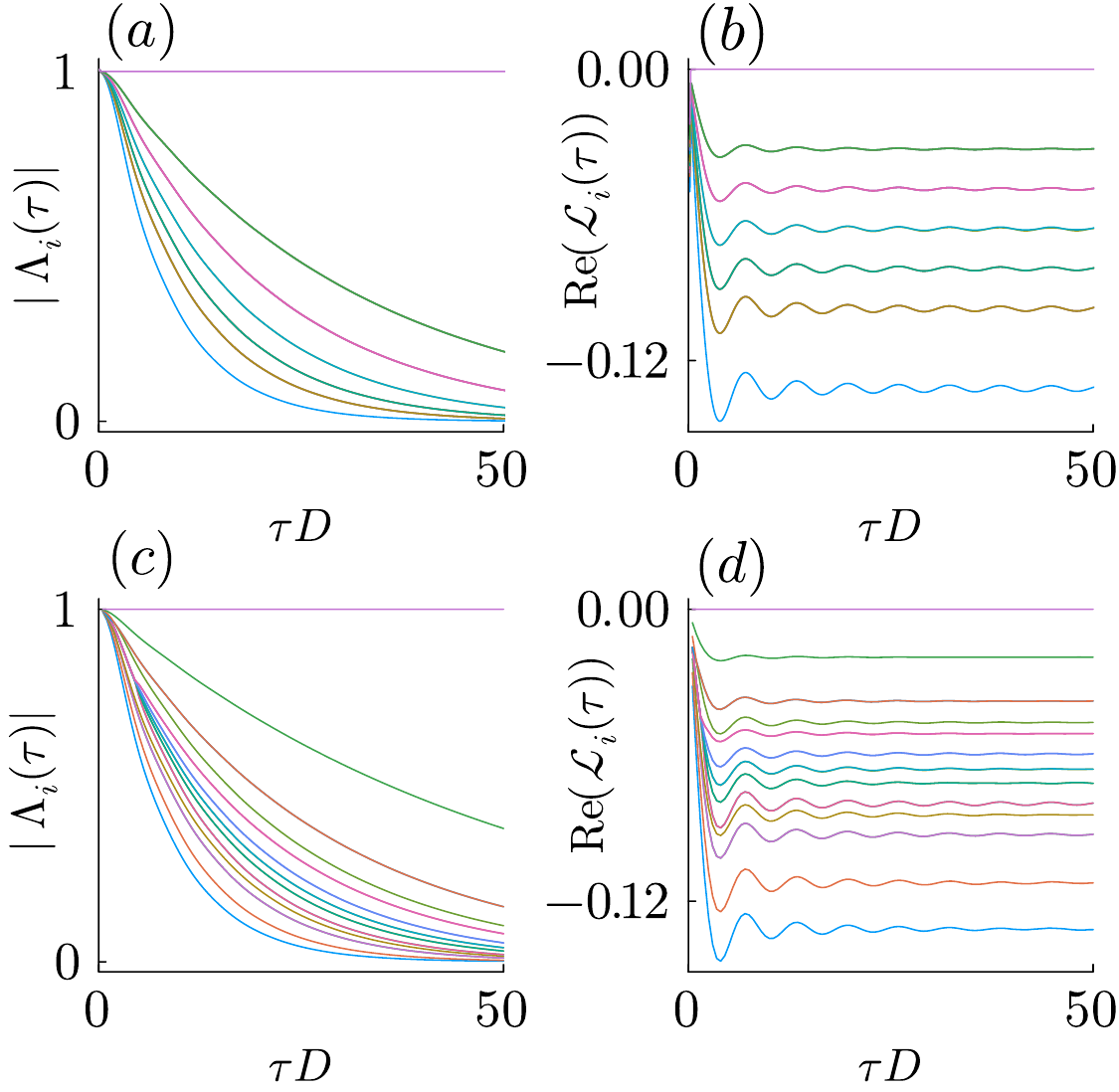}
  \caption{The eigenvalues $|\Lambda_{i}(\tau)|$ of the dynamical map $\hat{\Lambda}(\tau)$ and $\textrm{Re}(\mathcal{L}_{i}(\tau))$ of the propagator $\hat{\mathcal{L}}(\tau)$ shown against the time $\tau$. (a),(b) show the non-interacting case and (c),(d) show the interacting case $U=0.05D$. Parameters: $L = 3, t_{c}=0.02D, \Gamma_{\rm L} = \Gamma_{\rm R} = 0.05D, \beta_L=\beta_R = 1/D, \mu_L=-\mu_R = 0.2D.$}
  \label{fig: spectral plots}
\end{figure}

\section{Numerical results} \label{sec:results}
We now demonstrate the effectiveness of the dynamical map approach for paradigmatic fermionic systems. For the tensor network calculations, we use a time-step of $\delta \tau=0.1/D$, a 2TDVP truncation cutoff of $10^{-10} \sim 10^{-12}$ to time-evolve baths comprising of $M \sim 10^2$ modes and which generates bond dimensions on the order $\chi \sim 10^3$.

\textit{Spinless Fermi chain} - Consider a chain of spinless fermions with a Hamiltonian given by 
\begin{figure}[t]
  \centering
  \includegraphics[width=0.5\textwidth]{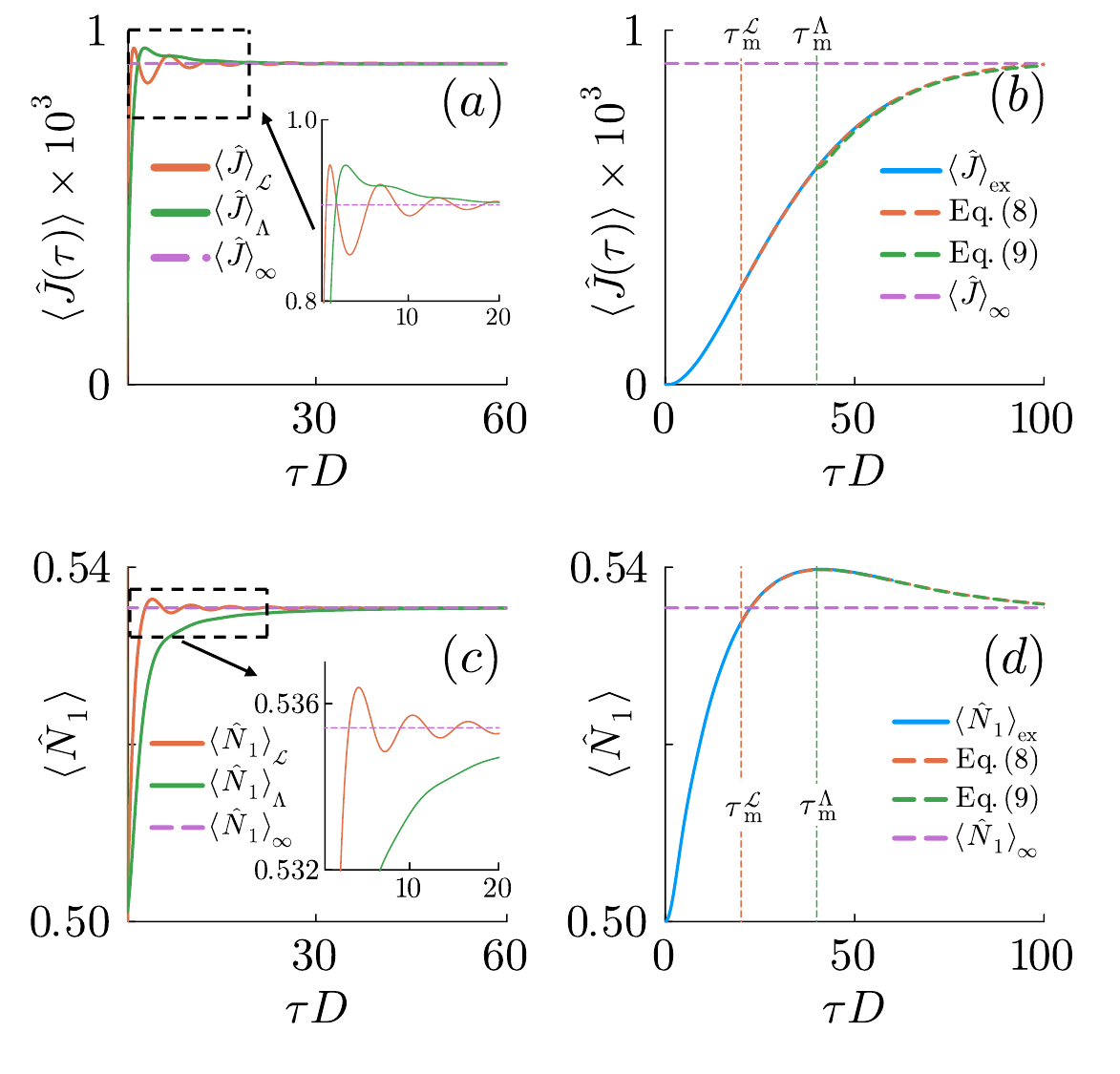}
  \caption{Evolution of a non-interacting Fermi chain in a non-equilibrium setup. (a) Convergence of the two time-dependent fixed point currents along with the exact steady state solution $\langle\hat{J}\rangle_{\infty}$, calculated using Landauer B\"uttiker theory. (b) The evolution of a totally mixed initial state (blue) is compared to the solution obtained from Eqs.~(\ref{eq:3}) and (\ref{eq:4}) using memory times $\tau^{\mathcal{L}}_{\rm m }D = 20$, $\tau^{\Lambda}_{\rm m}D = 40$. (c), (d) show the same analysis for the density on the first site. Parameters: $L = 3,U=0, t_{c}=0.02D, \Gamma_{\rm L} = \Gamma_{\rm R} = 0.05D, \beta_{\rm L}=\beta_{\rm R} = 1/D, \mu_{\rm L}=-\mu_{\rm R} = 0.2D.$}
  \label{non-interacting fermi chain}
\end{figure}
\begin{equation}
\hat{H}_{S} = \sum_{i=1}^{L-1}t_{c}(\hat{s}^{\dag}_{i+1}\hat{s}_{i}+ \hat{s}^{\dag}_{i}\hat{s}_{i+1})+U\hat{s}^{\dag}_{i}\hat{s}_{i}\hat{s}^{\dag}_{i+1}\hat{s}_{i+1},
\end{equation}
where each end of the chain is coupled to a bath, $\hat{Q}_{\rm L} = \hat{s}_{1},\hat{Q}_{\rm R} = \hat{s}_{L}$. We study the dynamics of a small $L=3$ site tight-binding chain with the system-bath couplings switched on at $\tau=0$. For simplicity, we assume a semi-elliptical spectral function for both baths
\begin{align} \label{eq:elliptical spectral function}
    \mathcal{J}_{\alpha}(\omega) &= \frac{2\Gamma_{\alpha}}{\pi^2}\sqrt{1-(\omega/D)^2},
\end{align}
with $\Gamma_{\rm L} = \Gamma_{\rm R}$ and $\beta_{\rm L}=\beta_{\rm R}$, but with $\mu_{\rm L}=-\mu_{\rm R}$ giving a non-equilibrium setup. 

We begin by considering the spectral properties of the map and propagator. In Fig.~\ref{fig: spectral plots}, the time-dependent eigenvalues $|\Lambda_{i}(\tau)|$ for $\hat{\Lambda}(\tau)$ and $\textrm{Re}(\mathcal{L}_{i}(\tau))$ for $\hat{\mathcal{L}}(\tau)$ are plotted for both non-interacting and interacting cases. They satisfy the properties of complete positivity outlined earlier in Sec.~\ref{sec:nonmarkov}. For $|\Lambda_{i}(\tau)|$ we see the fixed unit eigenvalue, corresponding to the time-dependent fixed point $\hat{\rho}^\Lambda_{\rm FP}(\tau)$, with the other eigenvalues decaying to zero with $\tau$ as $\hat{\Lambda}(\tau)$ converges to a projector onto the stationary state. For the decay rates $\textrm{Re}(\mathcal{L}_{i}(\tau))$ we also see a fixed zero eigenvalue corresponding to its time-dependent fixed point $\hat{\rho}^{\mathcal{L}}_{\rm FP}(\tau)$. The initial slippage at early times is clear in the transient behaviour of $\textrm{Re}(\mathcal{L}_{i}(\tau))$, after which they rapidly converge to constant values. The remaining oscillations observed correspond precisely to the bandwidth $D$ and hence occur on a much faster timescale than all other timescales of the problem. These are an artefact of the bath spectral function having a sharp band edge and disappear if this edge is smoothed over a larger bandwidth. Aside from degeneracies among the decay modes being lifted, we observe the same generic behaviour in Fig.~\ref{fig: spectral plots}(c)-(d) for the interacting case. Given the non-interacting results are computed exactly this serves as a first demonstration of our MPS methodology.

\begin{figure}[t]
  \centering
  \includegraphics[width=0.5\textwidth]{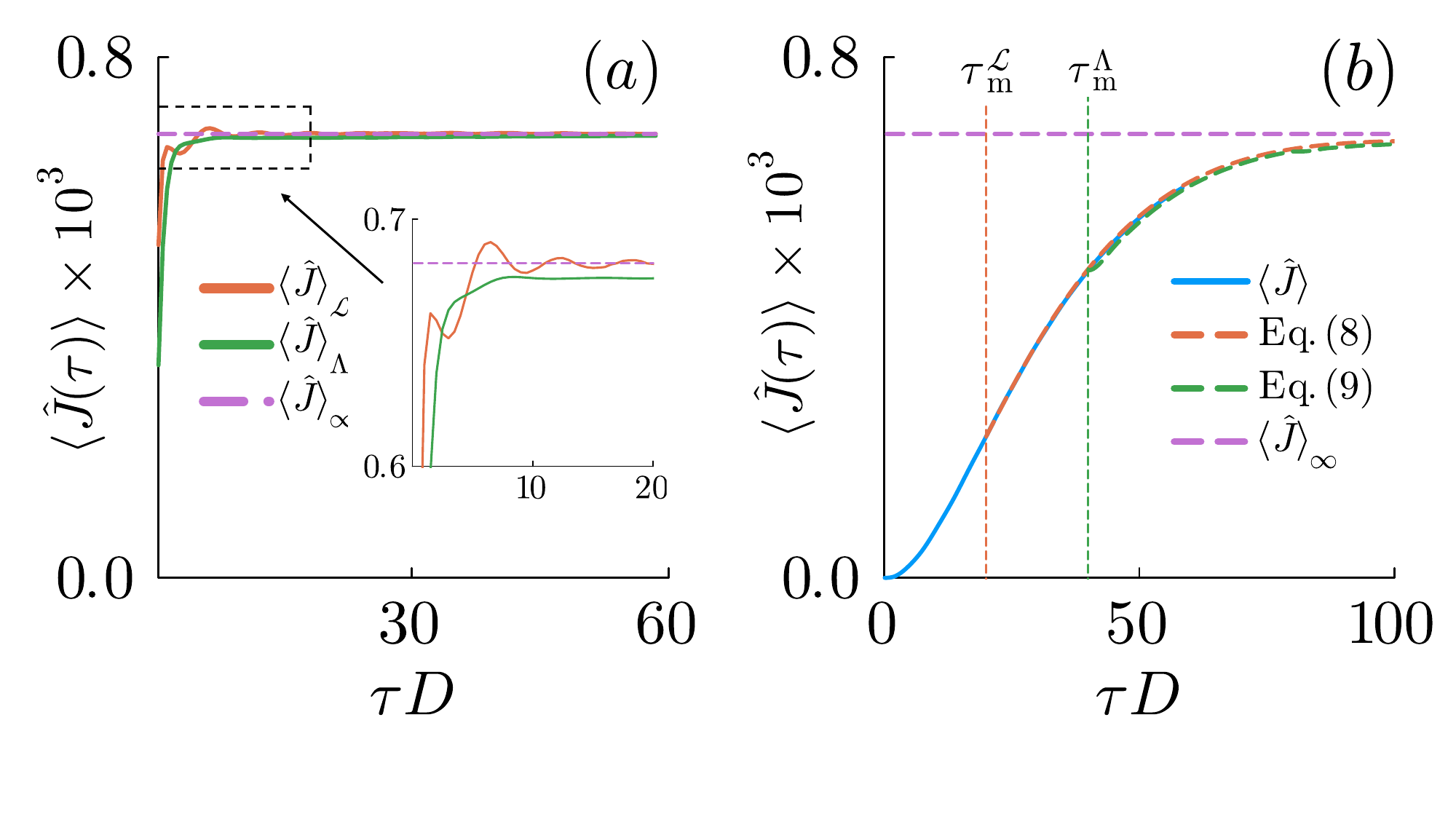}
  \caption{Evolution of an interacting Fermi chain in a non-equilibrium setup. (a) Convergence of the two time-dependent fixed point currents along with the steady state solution $\langle\hat{J}\rangle_{\infty}$ inferred from the largest $\tau$ calculated. (b) The evolution of a totally mixed initial state (blue) is compared to the predicted solution obtained from Eqs. (\ref{eq:3}) and (\ref{eq:4}) using memory times $\tau^{\mathcal{L}}_{\rm m }D = 20$, $\tau^{\Lambda}_{\rm m}D = 40$. Parameters: $L = 3, U=0.05D, t_{c}=0.02D, \Gamma_L = \Gamma_R = 0.05D, \beta_L=\beta_R = 1/D, \mu_L=-\mu_R = 0.2D.$}
  \label{interacting fermi chain}
\end{figure}

Next we consider some simple observables of the system, namely the density and the current, defined by $\hat{N}_{i} = \hat{s}^{\dag}_{i}\hat{s}_{i}$ and $\hat{J}_{i} = i(\hat{s}^{\dag}_{i}\hat{s}_{i+1} - \hat{s}^{\dag}_{i+1}\hat{s}_{i})$. The steady state current for non-interacting systems can be computed from Landauer-B\"{u}ttiker theory as
\begin{equation}
\langle\hat{J}_{\infty}\rangle = \frac{1}{2\pi}\int_{-D}^{D}{\rm d}\omega~\tau(\omega)[f_{\rm L}(\omega)-f_{\rm R}(\omega)],
\end{equation}
where $f_{\alpha}(\omega)$ denotes the Fermi-Dirac distribution for bath $\alpha$ and $\tau(\omega)$ is the transmission function. For details on the computation of $\tau(\omega)$, see Appendix~\ref{appendix: Landauer buttiker}. Figure~\ref{non-interacting fermi chain}(a) displays the evolution of the fixed point currents between site 2 and 3 of the system $\langle\hat{J}\rangle_{\mathcal{L}}\equiv \textrm{Tr}(\hat{J}_{2}\hat{\rho}^{\mathcal{L}}_{\rm FP}(\tau))$, $\langle\hat{J}\rangle_{\Lambda}\equiv \textrm{Tr}(\hat{J}_{2}\hat{\rho}^{\Lambda}_{\rm FP}(\tau))$ which show fast convergence to $\langle\hat{J}\rangle_{\infty}$ within $\tau\approx 20/D$. This is in contrast to the much longer thermalisation of a totally mixed initial state $\langle\hat{J}\rangle_{\rm ex}\equiv \textrm{Tr}(\hat{J}_{2}\hat{\rho}(\tau))$ shown in Fig.~\ref{non-interacting fermi chain}(b), which is still far from equilibrium by the end of the simulation at $\tau=60/D$. This demonstrates how our approach can offer a significant speedup in determining the stationary state compared to direct evolution.  

In addition to efficiently extracting steady states, we can extract all transient dynamics accurately using Eq.~\eqref{eq:3} once $\hat{\mathcal{L}}(\tau)$ has converged. As can be seen in Fig.~\ref{non-interacting fermi chain}(b), this approximation works remarkably well using $\tau^{\mathcal{L}}_{\rm m}D = 20$ recovering the exact dynamics up to $\tau=60/D$ and extrapolating beyond this demonstrating the steady state limit is attained. Applying the PreB approach of Eq.~\eqref{eq:4} with a larger memory time $\tau^{\Lambda}_{\rm m}D = 40$ recovers the transient dynamics reasonably well, but with a visible error expected due to the requirement that PReB should only be applied for times $\tau \gg \tau^{\Lambda}_{\rm m}$. Figures~\ref{non-interacting fermi chain}(c)-(d) show the same analysis for the density, where the hierarchy of timescales is again clear. The error in the PreB approach seen in Fig.~\ref{non-interacting fermi chain}(b) is several orders of magnitude smaller than the scale of Fig.~\ref{non-interacting fermi chain}(d) and so is not visible, showing both Eqs.~\eqref{eq:3} and \eqref{eq:4} recovering the correct transient dynamics and steady state limit. 

The non-interacting results reported were computed using the exact numerical solution outlined in Sec.~\ref{sec:noninteracting}. We have confirmed that these are fully reproduced by tensor network calculations (not shown). We now move to an identical setting with $U>0$ interacting chain, where tensor network methods are a necessity, and proceed with the same analysis. Figures~\ref{interacting fermi chain}(a)-(b) display the current with interactions identically to Figs.~\ref{non-interacting fermi chain}(a)-(b) for $U=0$. The value of the currents are suppressed due to interactions, but otherwise the behaviour is qualitatively identical. Specifically, there is a fast convergence for $\langle\hat{J}\rangle_{\mathcal{L}},\langle\hat{J}\rangle_{\Lambda}$ and a contrasting slow relaxation of $\langle\hat{J}\rangle_{\rm ex} \equiv\textrm{Tr}(\hat{J}_{2}\hat{\rho}(\tau))$. Again, the extrapolations using Eqs.~(\ref{eq:3})-(\ref{eq:4}) recover the transient dynamics accurately, with Eq.~(\ref{eq:4}) showing a small but visible slippage error. Together these results demonstrate that our dynamical map approach is viable for interacting Fermi systems. Our next example considers an seminal interacting problem with a well-known emergent timescale.

\begin{figure}[t!]
  \centering
  \includegraphics[width=0.5\textwidth]{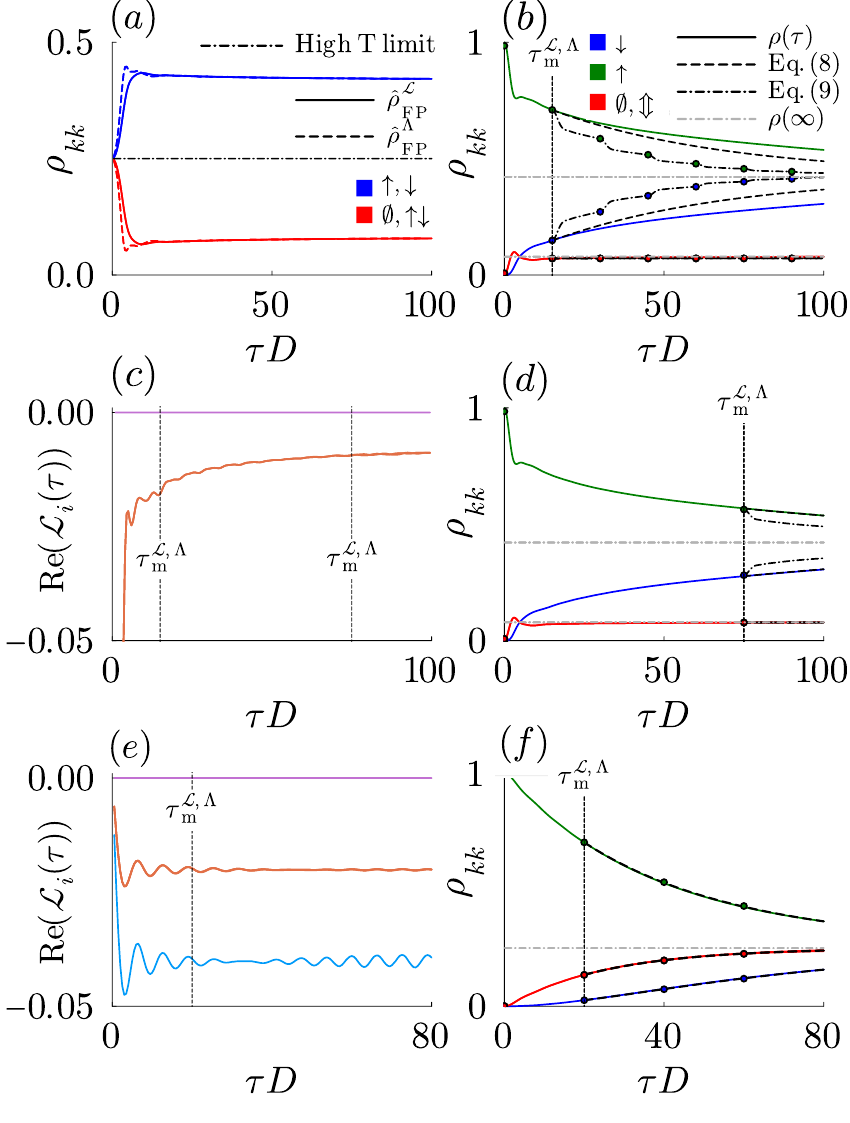}
  \caption{Evolution of the diagonal elements of the impurity density matrix $\hat{\rho}_{kk}$, where $k=\{\emptyset,\uparrow,\downarrow,\Updownarrow\}$ for the single impurity Anderson model. For strong interactions $U=4\Gamma$ at (i) moderate coupling $\Gamma=0.2D$ and low temperature $\beta=500/D$ Kondo physics is expected, while at (ii) weak coupling $\Gamma=0.02D$ and high-temperature $\beta=0.1/D$ the dynamics is expected to be more Markovian. (a) Convergence of the two time-dependent fixed points $\hat{\rho}^{\Lambda}_{\rm FP},\hat{\rho}^{\mathcal{L}}_{\rm FP}$ is shown for case (i). Note that due to particle-hole symmetry $\hat{\rho}_{\uparrow\uparrow} = \hat{\rho}_{\downarrow\downarrow},\hat{\rho}_{\emptyset\emptyset} = \hat{\rho}_{\Updownarrow\Updownarrow}$ for both fixed points. The totally mixed high-temperature fixed point for (ii) is also shown. 
 (b) The direct evolution of the initial state $\ket{\uparrow}\bra{\uparrow}$ for case (i) compared to the solution obtained from Eq.~(\ref{eq:3}) (dashed) and Eq.~(\ref{eq:4}) (dot-dashed) using memory times $\tau^{\mathcal{L}}_{\rm m}D = 15$, $\tau^{\Lambda}_{\rm m}D = 15$, respectively. The result of repeated applications of $\hat{\Lambda}(\tau^{\Lambda}_{\rm m})$ on the initial state is plotted stroboscopically (circles). (c) The evolution of the decay modes $\textrm{Re}(\mathcal{L}_{i}(\tau))$ for the case (i), showing a long lived time dependence. (d) The same as (b) but a longer memory time  $\tau^{\mathcal{L}}_{\rm m}D = 75 = \tau^{\Lambda}_{\rm m}D$ is used, where Eq.~(\ref{eq:3}) now accurately extrapolates the correct time evolution. (e) The evolution of the decay modes for the case (ii), showing convergence to fixed points up to bandwidth oscillation. (f) For case (ii) the direct solution of the initial state is again compared to the solution obtained from Eq.~(\ref{eq:3}) (dashed) obtained from using $\tau^{\mathcal{L}}_{\rm m }D = 20$. The solution using Eq.~(\ref{eq:4}) is only plotted stroboscopically (circles) as it's indistinguishable from Eq.~(\ref{eq:3}) for this case.}
  \label{EQ SIAM}
\end{figure}

\textit{Anderson impurity model} - We now consider the particle-hole symmetric single impurity Anderson model (SIAM), with Hamiltonian
\begin{equation}
\hat{H}_{S} = U\,\hat{s}^{\dag}_{\uparrow}\hat{s}_{\uparrow}\hat{s}^{\dag}_{\downarrow}\hat{s}_{\downarrow} -\frac{U}{2}\hat{s}^{\dag}_{\uparrow}\hat{s}_{\uparrow} -\frac{U}{2}\hat{s}^{\dag}_{\downarrow}\hat{s}_{\downarrow}.
\end{equation}
The Coulomb interaction $U\neq0$ gives rise to strong correlations both in and out of equilibrium causing complex behaviour such as the Kondo effect where the impurity entangles with the reservoir electrons screening its local moment~\cite{PhysRevB.91.085127,PhysRevB.88.094306,Coleman_2015}. The crossover between local moments and singlet formation is given by the Kondo temperature $T_K$, which for a single bath with spectral density $\mathcal{J}(\omega)$ is given by~\cite{Coleman_2015}
\begin{equation}
T_{K} = \sqrt{2U\mathcal{J}(0)}\,\textrm{exp}\left(-\frac{U}{8\mathcal{J}(0)}\right).
\end{equation}
Correspondingly the Kondo cloud bound state forms over a non-perturbatively long timescale $\tau_K \sim 1/T_K$.

To simulate the SIAM we treat it as a system of two spinless fermion modes, one for each spin projections $\sigma = \{\uparrow,\downarrow\}$, which are coupled only through the Coulomb interaction~\cite{kohn2021quenching}. Each spin projection $\sigma$ is then coupled via $\hat{Q}_\sigma = \hat{s}_\sigma$ to its own bath. We focus on the same equilibrium setup as in Ref.~\citen{PhysRevB.107.L201115}, so the $\sigma$ baths have identical spectral functions $\mathcal{J}(\omega)$, taken to be a flat spectral density with smoothed edges
\begin{align}
    \mathcal{J}(\omega) &= \frac{\Gamma}{2\pi(1+e^{\nu(\omega-D)})(1+e^{-\nu(\omega+D)})},
\end{align}
characterised by $\nu$ which we fix to be $\nu=100/D$, and identical coupling strength $\Gamma$, inverse temperature $\beta$ and chemical potential $\mu$. We initialise the impurity in the state $\hat{\rho}(t_0)=\ket{\uparrow}\bra{\uparrow}$ and fix the interaction to be strong as $U = 4\Gamma$.

The first scenario we consider is a moderately coupled bath\footnote{In Ref.~\cite{PhysRevB.107.L201115}, they use two identical baths and a coupling strength $\Gamma_{\rm L}=\Gamma_{\rm R}=D/10$ which has an equivalent total coupling strength as a single bath with $\Gamma=D/5$.} where the associated Kondo temperature is $T_{K} \approx 0.02D$ and with a bath temperature well below this. Being in the Kondo regime we expect the timescale $\tau_{K} = 50/D$ to manifest as a long memory time. In Fig.~\ref{EQ SIAM}(a) the time-dependent fixed points $\hat{\rho}^{\mathcal{L}}_{\rm FP},\hat{\rho}^{\Lambda}_{\rm FP}$ are shown to converge very rapidly by $\tau \approx 10/D$ to the thermal state suggesting a short memory time. In Fig.~\ref{EQ SIAM}(b) we recover and match the accuracy of the results from the influence tensor approach used in Ref.~\citen{PhysRevB.107.L201115} and similarly in Ref.~\citen{PhysRevB.107.125103} for the relaxation of all diagonal components of the impurities reduced density matrix. Even by $\tau=100/D$ the evolution of the spin polarised initial state is still far from equilibrium, reflecting the slow dynamical formation of a spin singlet in the strongly interacting Kondo regime expected for $\beta>1/T_K$. None of this physics is visible in the behaviour of the time-dependent fixed points but is instead revealed by the time dependence of the decay modes. In Fig.~\ref{EQ SIAM}(c) the decay modes fail to converge by $\tau=100/D$, showing a long-lived time dependence of the propagator. Note that only the slowest degenerate decay modes are shown as all other modes rapidly converge. This is reflected by the accuracy of the dynamical predictions given by Eqs.~(\ref{eq:3}) and (\ref{eq:4}). In Fig.~\ref{EQ SIAM}(b)
we see that when a small memory time $\tau_{\rm m}^{\mathcal{L}}$ is used, such that $\hat{\mathcal{L}}$ is far from converged, neither recover the transient dynamics. By increasing the memory time $\tau_{\rm m}^{\mathcal{L}}$ we can achieve accurate predictions as seen by the solution of Eq.~(\ref{eq:3}) in Fig.~\ref{EQ SIAM}(d), while Eq.~(\ref{eq:4}) still suffers visible slippage error. This points to the decay modes of $\hat{\mathcal{L}}(\tau),\hat{\Lambda}(\tau)$ having a strong dependence on system-bath correlations associated with the formation of the Kondo singlet screening cloud which is controlled by the much longer timescale $\tau_K$. Similar findings are reported in Refs.~\citen{PhysRevB.84.075150,PhysRevB.87.195108} when analysing the memory kernel in the Kondo regime.

In the second scenario we consider a weakly coupled bath where the associated Kondo temperature is $T_{K} \approx 0.001D$ and the bath temperature is far above this. In this high-temperature regime of a high-symmetry point the eventual thermal state is extremely close to the totally mixed state $\hat{\rho}_\infty =(\ket{\emptyset}\bra{\emptyset}+\ket{\uparrow}\bra{\uparrow}+\ket{\downarrow}\bra{\downarrow}+\ket{\Updownarrow}\bra{\Updownarrow})/4$. Correspondingly we find (not shown) that the time-dependent fixed points $\hat{\rho}_{\rm FP}^{\Lambda}(\tau),\hat{\rho}_{\rm FP}^{\mathcal{L}}(\tau)$ essentially immediately converge. While this is suggestive of a vanishingly short memory time, again, like with the Kondo regime, the predictive abilities and spectral properties of $\hat{\Lambda}(\tau)$ and $\hat{\mathcal{L}}(\tau)$ give a sharper characterisation. Figure~\ref{EQ SIAM}(e) displays the decay rates $\textrm{Re}(\mathcal{L}_{i}(\tau))$, showing an initial slippage before converging to fixed values up to bandwidth oscillations. The lack of a longer timescale is indicative that Kondo physics is not present. From this we take $\tau_{\rm m}^{\mathcal{L}}D = 20$ and see that Eq.~(\ref{eq:3}) accurately reproduces the true dynamics with minimal error, as shown in Fig.~\ref{EQ SIAM}(f). There is also strong agreement with Eq.~(\ref{eq:4}) whose stroboscopic predictions are displayed confirming that the physics of this regime can be fully captured within a very short time. We therefore see that the success or failure of the slippage approximations Eqs.~(\ref{eq:3})-(\ref{eq:4}) in capturing the transient dynamics serves as a very effective indicator of non-Markovian behaviour allowing the Kondo regime to be revealed.  

\section{Conclusions} \label{sec:conclusions}
Dynamical maps $\hat{\Lambda}(\tau)$ and time-local propagators $\hat{\mathcal L}(\tau)$ are central but often formal objects in the theory of open systems. Here we have shown that by combining the thermofield doubling and orthogonal polynomial chain mappings with the Choi-Jamiolkowski isomorphism both $\hat{\Lambda}(\tau)$ and $\hat{\mathcal L}(\tau)$ are readily accessible objects for numerical calculations of open Fermi systems. We work with pure state unitary evolution for which the numerical errors are well understood and positivity is manifestly preserved, in contrast to influence functional tensor methods\cite{PhysRevB.107.125103}. For systems which are mildly non-Markovian with a short memory time this dynamical map approach can accelerate the computation of stationary properties compared to directly simulating relaxation in the long-time limit. Moreover, converged $\hat{\Lambda}(\tau)$ and $\hat{\mathcal L}(\tau)$ allow for predictions of the transient dynamics for all longer times up to stationarity fully characterising the open system. We have demonstrated this for a spinless Fermi chain driven out of equilibrium by two baths, both with and without interactions, confirming the viability and effectiveness of this approach. We also investigated the equilibration of the SIAM hosting more complex Kondo physics. Here the thermal stationary state, even at strong coupling could be computed rapidly, but attempts to capture the transient relaxation dynamics were found to be challenging in the Kondo regime. Consequently this provides a robust means of quantifying the memory time and revealing non-trivial emergent effects known to distinguish different regimes of the SIAM. In the examples we reach the steady state rapidly, but there could be additional benefits to working directly in the steady state limit as done in Refs.~\citen{PhysRevLett.121.137702,PhysRevLett.130.186301}. The Choi-Jamiolkowski isomorphism is more general than the MPS methodology presented here and could also be combined with other methods such as the HEOM approach and process tensor approaches. Here we have focused exclusively on small systems comprising only $L=2-3$ spinless fermionic modes where both $\hat{\Lambda}(\tau)$ and $\hat{\mathcal L}(\tau)$ can be fully constructed as matrices. Future work\cite{Future_work} is looking into utilising tensor network methods for the entire calculations where both these objects are computed with a compressed description. This should enable much larger open systems to be addressed with a dynamical map approach. 

\begin{acknowledgments}
S.R.C. gratefully acknowledges financial support from UK's Engineering and Physical Sciences Research Council (EPSRC) under grant EP/T028424/1. A.P acknowledges funding from Seed Grant from IIT Hyderabad, Project No. SG/IITH/F331/2023-24/SG-169. A.P also acknowledges funding from Japan International Coorperation Agency (JICA) Friendship 2.0 Research Grant, and from Finnish Indian Consortia for Research and Education (FICORE).
\end{acknowledgments}

\section{Data Availability}
The data used to produced the plots is openly available at \href{https://doi.org/10.5523/bris.1qjxhltnvw1a2c0mtu4qvstwd}{https://doi.org/10.5523}.

\clearpage

\onecolumngrid

\setcounter{equation}{0}
\setcounter{figure}{0}
\setcounter{table}{0}

\renewcommand{\theequation}{S\arabic{equation}}
\renewcommand{\thefigure}{S\arabic{figure}}

\appendix 

\section{Finding the fermionic correction $\hat{P}$} \label{appendix: ancilla unitaries}
In terms of qubit operators, the unitary $\hat{P}$ is given by $ \hat{P} = \prod_{i=1}^{N_{s}}\hat{\sigma}^{x}_{a_{i}}$, where $\hat{\sigma}^{x}_{a_{i}}$ is the bit flip operator on the $i^{\textrm{th}}$ replica mode. In terms of fermionic operators, this is given by $\hat{P} = \prod_{i=1}^{L}\sum_{k=1}^{2i-1}e^{i\pi\hat{N}_{k}}(\hat{a}^{\dag}_{i} + \hat{a}_{i})$, where $k$ is summed over both system and replica modes. As each system mode is perfectly anti-correlated with its recplica mode initially, we can express any system number operator in terms of the corresponding replica's number operator, $\hat{N}_{k} = \mathbb{1}-\hat{N}_{k+1}$ where $k$ is even, such that $\hat{P}$ can be fully represented through replica modes. As the Hamiltonian only contains quadratic or quartic terms of system operators, we have $[\hat{P},\hat{H}] = 0$. If the interleaved mode ordering $\{\hat{d}_{i}\}_{\rm int}$ is used, the initial state is 
\begin{equation}
    \ket{\Psi^{\rm int}_{\rm CJ}} = \prod_{i=1}^{L}(\hat{s}^{\dag}_{i}+\hat{a}^{\dag}_{i})\ket{\mathbf{0},\mathbf{0},0,...0,1...,1}.
\end{equation}
We map $\ket{\Psi^{\rm int}_{\rm CJ}}$ to a state $\ket{\Psi_{\rm CJ}}$ using an additional unitary operation $\ket{\Psi_{\rm CJ}} = \hat{P}_{2}\ket{\Psi^{\rm int}_{\rm CJ}}$ which is given by a series of swap gates that are not fermionic. In practice, we don't swap the bath modes back into separated ordering for the isomorphism as they're all traced out as a contiguous block in either case.  In terms of fermionic operators, we have $\hat{P}_{2} = \prod_{i=1}^{L}\hat{T}_{i}^{q} = \prod_{i=1}^{L}e^{-i\pi\hat{N}_{i}\hat{N}_{i+1}}\hat{T}_{i}^{f}$ where $T_{i}^{q},\hat{T}_{i}^{f}$ are qubit and fermionic swap gates respectively, acting on modes $i,i+1$ where $\hat{N}_{i}$ is the number operator for mode $i$. These swaps will be between a replica and a system mode, or two replica modes. In either case, the phase correction can be expressed in terms of only replica modes due to the perfect correlation between $\hat{s}_{i}$ and $\hat{a}_{i}$ such that $[\hat{P}_{2},\hat{H}] = 0$. 

\section{Landauer B\"{u}ttiker Theory} \label{appendix: Landauer buttiker}

Here we briefly outline how to compute the currents in Landauer-B\"{u}ttiker theory which act as our point of comparison for non-interacting system steady states. In the continuum limit for macroscopic baths, the particle and energy currents in the absence of system interactions are given by \cite{mrsbulletin_2017,Purkayastha_2019}
\begin{equation}  \label{eq:23}
J_{LB}^{P} = \frac{1}{2\pi}\int_{-D}^{D}d\omega\tau(\omega)[f_{L}(\omega)-f_{R}(\omega)],
\end{equation}
and
\begin{equation}  \label{eq:24}
J_{LB}^{E} = \frac{1}{2\pi}\int_{-D}^{D}d\omega\omega\tau(\omega)[f_{L}(\omega)-f_{R}(\omega)],
\end{equation}
where $f_{\alpha}(\omega)$ denotes the Fermi-Dirac distribution for bath $\alpha$ and $\tau(\omega)$ is the transmission function of the system. This can be calculated in terms of the non-equilibrium Green's function \cite{PhysRevB.89.165105,Purkayastha_2019}. In our case this is given by 
\begin{equation}  \label{eq:25}
G(\omega) = \boldsymbol{M}^{-1}(\omega),
\end{equation}
with
\begin{equation}  \label{eq:26}
 \boldsymbol{M}(\omega) = \omega\mathds{1}-\boldsymbol{h}_{S}-\boldsymbol{\Sigma}^{(1)}(\omega) - \boldsymbol{\Sigma}^{(L)}(\omega),
\end{equation}
where the only nonzero elements of the self-energy matrices of the leads $\boldsymbol{\Sigma}^{(j)}(\omega)$ are 
\begin{equation}  \label{eq:27}
[\boldsymbol{\Sigma}^{(j)}]_{jj}(\omega) = \mathcal{P}\int_{-D}^{D}d\omega'\frac{\mathcal{J}_{j}(\omega')}{\omega'-\omega} - i\pi\mathcal{J}_{j}(\omega),
\end{equation}
using $\mathcal{J}_{1}(\omega) = \mathcal{J}_{L}(\omega)$, $\mathcal{J}_{L}(\omega) = \mathcal{J}_{R}(\omega)$ and $\boldsymbol{h}_{S}$ is defined via $\hat{H}_{S} = \sum_{i,j=1}^{L}(\boldsymbol{h}_{S})_{ij}\hat{s}^{\dag}_{i}\hat{s}_{j}$.
If the system Hamiltonian is of the form 
\begin{equation}  \label{eq:28}
\boldsymbol{h}_{S} = \sum_{j=1}^{L}\epsilon_{j}\hat{s}^{\dag}_{j}\hat{s}_{j} +\sum_{j=1}^{L-1}t_{i}(\hat{s}^{\dag}_{j+1}\hat{s}_{j} +\text{h.c.}),
\end{equation}
the transmission function is given by

\begin{equation}  \label{eq:29}
\tau(\omega) = 4\pi^{2}\mathcal{J}_{L}(\omega)\mathcal{J}_{R}(\omega)|[G(\omega)]_{1D}|^{2} 
= \frac{\mathcal{J}_{L}(\omega)\mathcal{J}_{R}(\omega)}{|\text{det}[\boldsymbol{M}]|^2}\prod_{i=1}^{
L-1}|t_{i}|^2.
\end{equation}

\section{Correlation matrix propagation} \label{appendix: corr propagation}
Here we prove Eq.~(\ref{eq:20}). Defining $\mathcal{U}(\tau) = e^{-i\hat{H}\tau}$, we have
\begin{align} \label{eq:58}
\boldsymbol{C}_{ij}(\tau) &= \bra{\psi(\tau)}\hat{d}^{\dag}_{j}\hat{d}_{i}\ket{\psi(\tau)} \nonumber = \bra{\psi(0)}\mathcal{U}^{\dag}(\tau)\hat{d}^{\dag}_{j}\hat{d}_{i}\mathcal{U}(\tau)\ket{\psi(0)}.
\end{align}
To evaluate this, first consider how $\hat{H}$ acts on $\hat{d}^{\dag}_{j}$,
\begin{align} 
\hat{H}\hat{d}^{\dag}_{j} &= \sum_{kl}\boldsymbol{H}_{kl}\hat{d}^{\dag}_{k}\hat{d}_{l}\hat{d}^{\dag}_{j} = \sum_{kl}\boldsymbol{H}_{kl}\hat{d}^{\dag}_{k}(\delta_{lj} - \hat{d}^{\dag}_{j}\hat{d}_{l}) = \sum_{k}\boldsymbol{H}_{kj}\hat{d}^{\dag}_{k}.
\end{align}
We can now evaluate $\mathcal{U}^{\dag}\hat{d}^{\dag}_{j}$ as
\begin{align}  \label{eq:60}
\mathcal{U}^{\dag}\hat{d}^{\dag}_{j} &= e^{iHt}\hat{d}^{\dag}_{j} = e^{i\sum_{k}\boldsymbol{H}_{kj}t}\hat{d}^{\dag}_{k} = \sum_{k}\boldsymbol{U}^{\dag}_{kj}\hat{d}^{\dag}_{k},
\end{align}
Substituting this into Eq.~(\ref{eq:58}) gives the result
\begin{align}  \label{eq:61}
\boldsymbol{C}_{ij}(t) &= \bra{\psi(0)}\sum_{kl} \boldsymbol{U}^\dagger_{kj}\hat{d}^\dagger_{k}(\boldsymbol{U}^\dagger_{li}\hat{d}^\dagger_{l})^\dagger \ket{\psi(0)} = \sum_{kl} \boldsymbol{U}^\dagger_{kj}\boldsymbol{U}_{il}\bra{\psi(0)}\hat{d}^{\dagger}_{k}\hat{d}_{l}\ket{\psi(0)} = \sum_{kl} \boldsymbol{U}^\dagger_{kj}\boldsymbol{U}_{il} \boldsymbol{C}_{lk}(0).
\end{align}
Thus we have
\begin{equation}  \label{eq:62}
\boldsymbol{C}(t) = \boldsymbol{U}\boldsymbol{C}(0)\boldsymbol{U}^{\dag}.
\end{equation}

\bibliography{Methodology_references.bib}

\end{document}